\documentclass[11pt,a4paper]{article}
\usepackage{jheppub}
\usepackage{./jheppub}
\usepackage[utf8]{inputenc}
\usepackage[english]{babel}
\usepackage[T1]{fontenc}
\usepackage[titletoc]{appendix}
\usepackage{titletoc}
\usepackage{graphicx}
\usepackage{empheq}
\usepackage{subcaption} 
\usepackage{xcolor}
\usepackage{amsmath,amssymb,amsthm}
\usepackage{bm}
\usepackage{float}
\usepackage[format=plain,labelfont={bf,it},textfont=it]{caption}
\usepackage{multirow}
\usepackage{slashed} 
\usepackage{tabularx}
\usepackage{bbold}
\usepackage[shortlabels]{enumitem}
\usepackage{tikz}
\usetikzlibrary{positioning,arrows,calc}
\usetikzlibrary{arrows.meta}
\usetikzlibrary{decorations.pathmorphing}
\usetikzlibrary{decorations.markings}
\usetikzlibrary{shapes.geometric}
\usetikzlibrary{patterns}
\tikzset{
  snake it/.style={
    decorate, 
    decoration=snake,
    segment length=3
  }
}
\usepackage{xparse}
\usepackage{pdfpages} 
\usepackage{titlesec} 
\usepackage{fancyhdr} 
\usepackage{emptypage} 
\usepackage{dsfont} 

\usepackage{indentfirst}
\usepackage{yfonts}
\usepackage{pgfplots}
\pgfplotsset{compat=1.18}
\usepgfplotslibrary{fillbetween}
\usetikzlibrary{patterns, positioning, fadings, arrows.meta, calc,decorations.text}
\hypersetup{
  pdftitle={Regge in Wonderland},
  pdfauthor={Alessio Miscioscia, Dmitrii Pavshinkin, Fedor K. Popov},
  pdfsubject={Large-spin conformal field theory on pp-wave backgrounds},
  pdfkeywords={Scale and Conformal Symmetries, Thermal Field Theory, Effective Field Theories, 1/N Expansion}
}
\title{Regge in Wonderland}
\preprint{YITP-SB-2026-15}

\author[a]{Alessio Miscioscia,}
\author[a,b]{Dmitrii Pavshinkin}
\author[b]{and Fedor K. Popov}

\affiliation[a]{C. N. Yang Institute for Theoretical Physics, Stony Brook University, Stony Brook, NY 11794,USA}
\affiliation[b]{Simons Center for Geometry and Physics, Stony Brook University, Stony Brook, NY 11794,USA}

\emailAdd{alessio.miscioscia@stonybrook.edu}
\emailAdd{dmitrii.pavshinkin@stonybrook.edu}
\emailAdd{fedor.popov@scgp.stonybrook.edu}

\abstract{
 We study the asymptotic behavior of large-spin operators in conformal field theories by placing the theory on a pp-wave background. In this setting, the diagonal three-point coefficients involving two large-spin operators and a light operator are mapped to thermal one-point functions on the pp-wave. We predict these one-point functions in the two complementary regimes of high and low temperature, corresponding respectively to the total twist parametrically above and below an appropriate spin-dependent scale. In particular, at high temperature, we derive their asymptotic behavior using thermal effective field theory, while at low temperature, we obtain universal formulas that can be matched directly to predictions from the analytic bootstrap. We test these results in free theories on the pp-wave. Finally, we solve the large-$N$ $\mathrm{O}(N)$ model numerically, finding agreement with our predictions for both the one-point function and the free energy density in the high- and low-temperature regimes.
}

\begin{document}

\maketitle
\section{Introduction and Summary}
Conformal field theories (CFTs) play a central role in quantum field theory as they describe renormalization-group fixed points and thereby govern continuous phase transitions and scale-invariant regimes.
The conformal data of a CFT consist of the spectrum of local primary operators and the operator product expansion (OPE) coefficients. Each primary is labeled by its scaling dimension, Lorentz representation, and global-symmetry quantum numbers. These data must satisfy the constraints imposed by locality, unitarity, and crossing symmetry. 
Understanding these constraints is one of the central problems in the subject.

Substantial progress has been achieved in several complementary regimes of the CFT spectrum. Numerical conformal bootstrap methods can place bounds on scaling dimensions and OPE coefficients for operators with quantum numbers of order one \cite{Poland:2018epd}. On the other hand, analytic results are also available for very heavy operators. Thus, in two dimensions, the Cardy formula determines the asymptotic density of states at large scaling dimension \cite{Cardy:1986ie}. Analogous high-energy asymptotics in higher-dimensional CFT follow from the thermal partition function and the locality of the thermal effective action \cite{Shaghoulian:2015kta,Benjamin:2023qsc,Benjamin:2024kdg,Banerjee:2012iz} (see also \cite{Simmons-Duffin:2025qox,Buric:2025uqt,Buric:2026pes} for the asymptotics of three-point functions). Another regime accessible to analytic methods is associated with the large global charge. Here, the low-lying spectrum at fixed charge can be organized by an effective field theory \cite{Monin:2016jmo,Cuomo:2021cnb,Cuomo:2022kio,Hellerman:2015nra}.

In the light-cone bootstrap, or conformal Regge theory, one studies operators whose spin becomes infinitely large while their twist remains fixed \cite{vanRees:2024xkb,Alday:2007mf}. Multi-twist operators in this regime can be described in terms of weakly interacting constituents \cite{Komargodski:2012ek}, and their anomalous dimensions can be computed systematically in the large-spin expansion \cite{Alday:2015eya,Fitzpatrick:2012yx,Caron-Huot:2017vep,Fardelli:2025fkn,Fardelli:2025eun,Simmons-Duffin:2016wlq}. As the twist increases, the interactions grow stronger, and this semiclassical approximation eventually breaks down.
In $2+1$ dimensions, the interactions among the effective constituents become  comparable to the leading contribution to the twist at the characteristic
scale \cite{Kravchuk:2024wmv,SemiUniversality,Komargodski:2026ain}
\begin{equation}
    \tau \sim \sqrt{J_z}\,,
    \qquad \tau=\Delta-J_z.
\end{equation}
In $3+1$ dimensions, when both angular momenta are large with fixed
ratio, the analogous strongly interacting regime is reached for
\cite{SemiUniversality,Komargodski:2026ain}
\begin{equation}
    \tau \sim (J_1J_2)^{1/3}\,,
    \qquad \tau=\Delta-J_1-J_2.
\end{equation}
These scales mark a crossover between the weakly-coupled Regge regime and a genuinely strongly coupled sector of fast-spinning operators.
For recent developments, see
\cite{Lee:2026azy,Mukherjee:2026dfu,Advant:2026cqt}%
\footnote{We refer to \cite{Deb:2025ypl,Deb:2025ddc} for similar limits in super-conformal field theories.}.

To study this crossover, 
one takes correspondingly the double-scaling limits in $(2+1)d$
\begin{equation}
    J_z\to\infty,
    \qquad
    \frac{\tau}{\sqrt{J_z}}=\text{fixed},
\end{equation}
and in $(3+1)d$
\begin{equation}
    J_1,J_2\to\infty,
    \qquad
    \frac{J_1}{J_2}=\text{fixed},
    \qquad
    \frac{\tau}{(J_1J_2)^{1/3}}=\text{fixed}.
\end{equation}
 Recent work showed that these limits admit a geometric description: after an appropriate null limit of the Lorentzian cylinder, the CFT is naturally formulated on an effective pp-wave background \cite{Komargodski:2026ain}. The emergent Heisenberg symmetries of these backgrounds, combined with locality, explain the universal extensive form of the large-spin partition function and constrain the corresponding asymptotic density of states\footnote{Geometries similar to the pp-wave discussed in this paper also appeared previously in the literature in different contexts \cite{Blau:PlaneWavesPenroseLimits,Metsaev:2002re,BMN2002,Maldacena:2008w}. The holographic dual of the story was studied in  \cite{Dorey:2022cfn,Dorey:2023jfw,Mouland:2023gcp}.}.

Although the pp-wave partition function has been shown to determine the asymptotic density of large-spin states,
much less is known about the remaining conformal data in this regime. In particular, the heavy-light-heavy OPE coefficients are needed to characterize the dynamics of large-spin states beyond their degeneracies. By the state-operator correspondence, diagonal heavy-light-heavy (HLH) coefficients can be extracted from one-point functions of light operators in the corresponding heavy states. The pp-wave description suggests a natural route to their computation. One can evaluate one-point functions of local operators in the effective pp-wave geometry and translate the result back into asymptotic OPE data of the original CFT. The purpose of this work is to develop this approach and use it to probe correlation functions in the large-spin regime.

When only one angular momentum is taken to be large, it is sufficient to consider the case of a $(2+1)$-dimensional CFT. The generalization to higher dimensions is analogous. The corresponding pp-wave geometry is 
\begin{equation}
ds_{\rm pp}^2=-(1+y^2)\,dt^2+2\,dt\,dx+dy^2.
\end{equation}
The isometries constrain thermal one-point functions on this background to be functions only of $\beta=1/T$ and $y$. Although their general form is theory dependent, both the high- and low-temperature limits can be studied analytically.

In the high-temperature limit, $\beta\to0$, the one-point function must reduce locally to the infinite-volume flat-space result. For a parity-even scalar probe, this statement follows from the
local thermal effective field theory expansion, under the assumptions
discussed below, along the lines of \cite{Benjamin:2023qsc,Benjamin:2024kdg,Diatlyk:2024qpr}. For a non-zero flat-space thermal one-point coefficient $b_{\mathcal O}$, we find\footnote{Those coefficients are in principle related to the conformal data, but it is usually more convenient to use alternative paths to compute them non-perturbatively \cite{Iliesiu:2018fao,Iliesiu:2018zlz,Marchetto:2023xap,Barrat:2025wbi,Barrat:2025nvu}.}
\begin{equation}
    \langle \mathcal O(y)\rangle_{{\rm pp},\beta}
    \overset{\beta\to0}{\simeq}
    \frac{b_{\mathcal O}}
    {\beta^{\Delta_{\mathcal O}}(1+y^2)^{{\Delta_{\mathcal O}}/2}}
    \left[1+O(\beta^2)\right],
\end{equation}
where the position dependence comes from the gravitational redshift $\beta\sqrt{-g_{tt}(y)}$.

After a Weyl rescaling and inverse Laplace transformation of this one-point function, one determines the averaged diagonal matrix elements on the cylinder $\mathbb{R}\times S^2$ in the  large-twist regime,
\begin{equation}
\left\langle \tau,J_z|  \mathcal{O}(\theta) \right| \tau,J_z\rangle
 \simeq \frac{b_{\mathcal O}(\tau J_z)^{\Delta_{\mathcal O}/6}}{\left[1+\frac{J_z}{2\tau}\left(\theta-\frac{\pi}{2}\right)^2\right]^{\Delta_{\mathcal O}/2}},\qquad \sqrt{J_z}\ll\tau\ll J_z,
\end{equation}
where $\theta$ denotes the angle on $S^2$.

However, the low-temperature regime on the pp-wave is dominated by the light spectrum of the theory. Keeping the leading contribution of a scalar primary $\mathcal P$ with a non-vanishing OPE coefficient $C_{\mathcal P\mathcal P\mathcal O}$, normalization $C_\mathcal{P}$, and twist $\tau_\mathcal{P}$, we find
\begin{empheq}{multline}
\langle\mathcal O(y)\rangle_{{\rm pp},\beta}
\overset{\beta\to\infty}{\simeq}
e^{-\beta\tau_{\mathcal P}}
\left(\frac{2}{\beta}\right)^{\Delta_{\mathcal O}/2}
\frac{C_{\mathcal P\mathcal P\mathcal O}}{C_{\mathcal P}}
\frac{
\Gamma(\tau_{\mathcal P})
\Gamma(\Delta_{\mathcal O}-1)
}{
\Gamma(\Delta_{\mathcal O}/2)
\Gamma(\tau_{\mathcal P}+\Delta_{\mathcal O}/2-1)
}
\\
\times
{}_2F_1\!\left(
\Delta_{\mathcal O}-1,
\frac{\Delta_{\mathcal O}}{2};
\tau_{\mathcal P}+\frac{\Delta_{\mathcal O}}{2}-1;
-\frac{2y^2}{\beta}
\right),\qquad \Delta_{\mathcal O}>1.
\label{eq:intro-1f-1pt-pp}
\end{empheq}
In a special case of this formula, when $\Delta_{\mathcal{O}}=2\tau_{\mathcal{P}}$, the hypergeometric function reduces to a simple power law,
\begin{equation}
\label{eq:intro-special-one-point}
\langle
\mathcal O(\vec y)
\rangle_{\rm pp,\beta}
\overset{\beta\to\infty}{\simeq}\frac{
C_{\mathcal P\mathcal P\mathcal O}
}{
C_{\mathcal P}}\frac{e^{-\beta\tau_{\mathcal P}}}{(\beta/2+|\vec{y}|^2)^{\tau_{\mathcal P}}},\qquad \tau_{\mathcal{P}}\geq \frac12.
\end{equation}

Converting the low-temperature response into the corresponding HLH correlator on a cylinder requires distinguishing the lightest primary $\mathcal{P}$ that couples diagonally to the probe $\mathcal{O}$ and the minimal twist primary with $\tau_{\min}\leq\tau_\mathcal{P}$ that controls the thermodynamic ensemble \cite{Komargodski:2026ain},
\begin{equation}
\mathcal{Z}={\rm Tr}e^{-\beta\tau-\mu J_z},\qquad
\log\mathcal{Z}\sim \frac{e^{-\beta\tau_{\min}}}{\mu},\qquad \beta\to \infty.
\end{equation}
In the parametrically small twist regime, $1\ll\tau\ll\sqrt{J_z}$, the inverse transform is dominated by the saddle
\begin{equation}
\beta_*
\simeq
\frac{2}{\tau_{\min}}
\log\!\left(
\frac{\tau_{\min}\sqrt J_z}{\tau}
\right),
\qquad
\mu_*
\simeq
\frac{\tau}{\tau_{\min}J_z},
\end{equation}
and the asymptotic scaling becomes
\begin{multline}
\label{eq:asymptotics2+1}
\left\langle \tau,J_z \middle| \mathcal{O}(\theta) \middle| \tau,J_z \right\rangle
\sim
\left({\tau_{\min}}/{\tau}\right)^{
\Delta_{\mathcal O}/2-2\tau_{\mathcal P}/\tau_{\min}}
J_z^{\Delta_{\mathcal O}/2-\tau_{\mathcal P}/\tau_{\min}}
\\
\times {}_2F_1\!\left(
\Delta_{\mathcal O}-1,
\frac{\Delta_{\mathcal O}}{2};
\tau_{\mathcal P}+\frac{\Delta_{\mathcal O}}{2}-1;
-\frac{\tau_{\min}J_z}{\tau}
\left(\theta-\frac{\pi}{2}\right)^2
\right),
\qquad
\tau\ll\sqrt{J_z}.
\end{multline}

The case of $(3+1)d$ CFTs on $\mathbb{R}\times S^3$ with two large momenta $J_1$ and $J_2$ and fixed $J_1/J_2$ is even more constrained. The corresponding pp-wave metric is
\begin{equation}
    ds_{\rm pp_2}^2=-dt^2-2\,dt\,dv-4u\,dt\,dy+du^2+dy^2.
\end{equation}
The isometries of this metric force the thermal one-point functions of scalar operators to be constant in space and to depend only on $\beta$. Their finite-temperature form is again theory dependent, but the two extreme limits are under analytic control. At high temperature, we find
\begin{equation}
    \langle\mathcal O\rangle_{{\rm pp_2},\beta}
    \overset{\beta\to0}{\simeq}
    \frac{b_{\mathcal O}}{\beta^{\Delta_{\mathcal O}}}
    \left[1+O(\beta^2)\right].
\end{equation}
The low-temperature limit is again controlled by the light spectrum and can be computed explicitly. The low- and high-temperature limits correspond, respectively, to
\begin{equation}
   1\ll{\tau}\ll(J_1J_2)^{1/3}\;\;\text{and}\;\;
    (J_1J_2)^{1/3}\ll\tau\ll\min\{J_1,J_2\}.
\end{equation}

After the Weyl rescaling and inverse Laplace transform, the averaged HLH correlator on the cylinder has the form
\begin{equation}
\label{eq:asymptotics3+1}
\left\langle
\tau,J_1,J_2
\middle|
\mathcal{O}(\vartheta)
\middle|
\tau,J_1,J_2
\right\rangle
\sim
\begin{cases}
\displaystyle
\frac{
\left({\tau_{\min}}/{\tau}\right)^{
\Delta_{\mathcal O}/2-3\tau_{\mathcal P}/\tau_{\min}}
(J_1J_2)^{
\Delta_{\mathcal O}/2-\tau_{\mathcal P}/\tau_{\min}}
}{
\left(J_1\cos^2\vartheta+J_2\sin^2\vartheta\right)^{
\Delta_{\mathcal O}/2}
},
&
\tau\ll(J_1J_2)^{1/3},
\\[1.9em]
\displaystyle
\frac{b_{\mathcal O}
(\tau J_1J_2)^{\Delta_{\mathcal O}/4}
}{
\left(J_1\cos^2\vartheta+J_2\sin^2\vartheta\right)^{
\Delta_{\mathcal O}/2}
},
&
\tau\gg(J_1J_2)^{1/3},
\end{cases}
\end{equation}
where $0\leq\vartheta\leq\pi/2$ parametrizes the position on $S^3$ relative to the two orthogonal rotation planes associated with $J_1$ and $J_2$, and the light scalar primaries have the twists $\tau_\mathcal{P}\geq\tau_{\rm min}\geq1$.

Finally, we consider thermal one-point functions of pseudoscalar operators on the $(2+1)d$ pp-wave. For a pseudoscalar primary $\mathcal{O}$ of dimension $\Delta_\mathcal{O}$, parity forbids a homogeneous thermal expectation value, while the vorticity pseudoscalar $\omega$ permits a contribution at first derivative order \cite{Jensen:2011xb,Manes:2013kka}. On the pp-wave background, the leading high-temperature response becomes
\begin{equation}
\langle{\mathcal O}(y)\rangle_{\mathrm{pp},\beta}
\overset{\beta\to 0}{\simeq}
\frac{b^{(\omega)}_{\mathcal O}}{\beta^{\Delta_{\mathcal O}-1}}
\frac{y}{(1+y^2)^{(1+\Delta_{\mathcal O})/2}} ,
\end{equation}
where, as in the scalar case, $b^{(\omega)}_{\mathcal O}$ is the dimensionless one-point function coefficient which can be determined either from conformal data or constrained by a proper thermal bootstrap problem \cite{Iliesiu:2018fao,David:2023uya}.

We test all our predictions in a number of explicit examples. In particular, we verify the results in free field theories in $2+1$ and $3+1$ dimensions. We also solve the critical large-$N$ $\mathrm{O}(N)$ model numerically on the pp-wave background. We compare the solution of the gap equation for the Hubbard--Stratonovich field with our high- and low-temperature predictions. As a byproduct we compute the pp-wave free energy of the model and compare it with the high- and low-temperature asymptotics given in \cite{Komargodski:2026ain}. We also test our low-temperature prediction using results from the analytic bootstrap. In the large-spin, low-twist regime, the theory admits a description in terms of weakly interacting partons on the cylinder.

The rest of this work is organized as follows. In Section \ref{eq:Seconepointppwave}, we derive general predictions for thermal one-point functions on the pp-wave background in both the high- and low-temperature regimes. Section \ref{sec:examples} presents explicit examples, including the free scalar and free fermion theories and the large-\(N\) \(\mathrm{O}(N)\) model, which we solve on the pp-wave\footnote{During the final stages of this work, we became aware of \cite{David:2026yis,mishra2026semiuniversalityonmodels1times}, which address the problem of the \(\mathrm{O}(N)\) model on the $(2+1)d$ pp-wave background using a different approach.}. Finally, in Section \ref{ref:HHLcylinder}, we compare our low-temperature result with the analytic-bootstrap description in terms of weakly interacting particles on the cylinder. The two appendices derive the pp-wave energy spectrum and show the conformal invariance of the pp-wave vacuum state.

\section{One-point functions on pp-wave}\label{eq:Seconepointppwave}
In \cite{Komargodski:2026ain}, the partition function of a conformal field theory is studied in the sector of operators carrying large angular momentum. It is shown that, in this limit, the dynamics is described by an emergent spacetime geometry, providing an effective description of the large-spin sector.

In $2+1$ dimensions, considering states with large eigenvalue of the angular momentum $J_z$ generating rotations in the spatial plane, the effective geometry is a Brinkmann pp-wave of the form
\begin{equation}
ds_{\rm pp}^2
=
-\left(1+y^2\right)dt^2
+
2\,dt\,dx
+
dy^2.
\end{equation}

In $3+1$ dimensions, there are two independent angular momenta, $J_1$ and $J_2$, associated with rotations in the $12$ and $34$ planes, respectively. Different scaling limits lead to different effective geometries. If only one angular momentum is taken to be large while the other is kept finite, the resulting geometry is the straightforward higher-dimensional generalization of the previous pp-wave,
\begin{equation}\label{eq:2largef}
ds_{\rm pp}^2
=
-\left(1+y_1^2+y_2^2\right)dt^2
+
2\,dt\,dx
+
dy_1^2
+
dy_2^2  .
\end{equation}
The generalization of this limit in any dimension is straightforward. 

On the other hand, if both angular momenta are taken to infinity while keeping their ratio
\(
J_1/J_2
\)
fixed, one obtains a different emergent geometry,
\begin{equation}
ds_{\rm pp_2}^2
=
-dt^2
-2\,dt\,dv
-4u\,dt\,dy
+
du^2
+
dy^2.
\end{equation}
These geometries capture the universal physics of the large-spin sector and provide an effective description of conformal field theories in the corresponding scaling limits.

\subsection{A single large angular momentum}

We begin by considering the pp-wave geometry with a single large angular momentum in $d\geq 3$ spacetime dimensions,
\begin{equation}
\label{eq:pp-wave1fug}
ds_{\rm pp}^2
=
-\left(1+|\vec{y}|^2\right)dt^2
+
2\,dt\,dx
+
\sum_{i=1}^{d-2}dy_i^2, \qquad |\vec{y}|^2=\sum\limits_{i=1}^{d-2}y_i^2.
\end{equation}
This metric is stationary and therefore admits the globally timelike Killing vector $K_0^\mu\partial_\mu=\partial_t$,  with $K_0^2=-(1+|\vec{y}|^2)$. Choosing this Killing flow as the time direction defines the corresponding conserved Hamiltonian $H_{\mathrm{pp}}$, and the thermal partition function becomes $\mathcal{Z}=\text{Tr}e^{-\beta H_{\mathrm{pp}}}$.

Expectation values of local operators are strongly constrained by the isometries of this background. At zero temperature, the symmetry group is generated by the $\mathbb H_{2d-3}$ Heisenberg algebra together with time translations, translations along the $x$ direction, and transverse rotations. The isometries constrain scalar one-point functions to be constant. Their vanishing for nonidentity scalar primary operators follows from the stronger assumption of a conformally invariant pp-wave vacuum; Appendix~\ref{ref:conformalinvariancePpwave} establishes this invariance in three dimensions for a lowest-energy state with vanishing null momentum.

At finite temperature, which is the regime of interest for our purposes,
the situation changes substantially. The thermal circle explicitly breaks the conformal symmetry relating the pp-wave geometry to flat space, and local operators generically acquire non-vanishing one-point functions. These expectation values depend on the temperature,
\(
T=1/\beta,
\)
as well as on the transverse coordinates through the invariant combination
\(|\vec{y}|
\),\footnote{The case of a parity-odd probe $\mathcal{O}$ in three spacetime dimensions is discussed separately in Section \ref{sec:pp-free-fermion}.}
\begin{equation}
\label{eq:one-point-T}
\langle\mathcal O(\vec{y})\rangle_{\mathrm{pp},\beta}
=
\mathcal{Z}^{-1}\operatorname{Tr}
\Big(
e^{-\beta H_{\mathrm{pp}}}
\mathcal O({\vec{y}})
\Big).
\end{equation}

Computing thermal one-point functions in an arbitrary conformal field theory is a difficult problem, and explicit results are available only in a limited number of examples. Nevertheless, two asymptotic regimes admit a universal description, which we shall discuss below.

\subsubsection{High-temperature limit}\label{sec:high-T}

We first consider the high-temperature regime $\beta\ll1$, and note that thermal equilibrium on a curved spacetime generally implies that different stationary observers measure different local temperatures. On the pp-wave background, the local inverse temperature is defined by the Tolman-Ehrenfest relation,\footnote{This is essentially the statement that the proper time of a local stationary observer is $d\tau=\sqrt{-g_{tt}}\,dt$, and, after Wick rotation, the circumference of the thermal circle becomes $\widetilde\beta=\int_0^\beta \sqrt{g_{00}}\,dt_E$.}
\begin{equation}
\widetilde\beta(\vec{y})
=
\beta\sqrt{-K_0^2}
=
\beta\sqrt{1+|\vec{y}|^2}.
\end{equation}

Taking the limit $\beta\to0$ at any fixed transverse position, the typical thermal wavelength becomes much smaller than the characteristic geometric scale. Consequently, local observers are insensitive to the global geometry of spacetime, and the thermal one-point functions are well approximated by the flat-space expectation values evaluated at the local inverse temperature,
\begin{equation}
\label{eq:high-T-one-point}
\langle\mathcal O(
\vec{y})\rangle_{\mathrm{pp},\beta}
\overset{\beta\to 0}{\simeq}
\langle\mathcal O\rangle_{\widetilde\beta}.
\end{equation}
 
 More generally, using the universal form of thermal one-point functions in flat space, we obtain the leading high-temperature expectation value of a symmetric traceless spin-$\ell$ primary on the pp-wave,
\begin{equation}\label{eq:highTexpected}
\langle
\mathcal O^{\mu_1\cdots\mu_\ell}(\vec{y})
\rangle_{\mathrm{pp},\beta}
\;\overset{\beta\to0}{\simeq}\;
\frac{
b_{\mathcal O}
}{
{\widetilde\beta}^{\Delta_{\mathcal O}}}
\left(
e^{\mu_1}\cdots e^{\mu_\ell}
-
\text{traces}
\right),
\end{equation}
where $b_{\mathcal O}$ is the thermal one-point coefficient of the corresponding operator in flat space and $e^\mu=K_0^\mu/\sqrt{-K_0^2}$ denotes the local unit timelike vector defining the thermal rest frame. In particular, applying this formula to the stress-energy tensor with $\Delta_{T}=d$, we find
\begin{equation}
\langle
T^{\mu\nu}(\vec{y})
\rangle_{\mathrm{pp},\beta}
\;\overset{\beta\to0}{\simeq}\;
\frac{
b_T
}{
\beta^{d}
(1+|\vec{y}|^2)^{{d}/2}
}
\left(g^{\mu\nu}+d\,e^\mu e^\nu
\right).
\end{equation}
This is exactly the stress-energy tensor of an ideal conformal fluid in hydrostatic equilibrium, 
\begin{equation}
\langle T^{\mu\nu}\rangle =(\varepsilon+p)e^\mu e^\nu+p  g^{\mu\nu},
\end{equation}
with local flat-space pressure $p(\vec{y})=b_T\widetilde\beta^{-d}$ and energy density $\varepsilon(\vec{y})=(d-1)p(\vec{y})$.

The argument above is independent of the precise form of the geometry \eqref{eq:pp-wave1fug} and can be generalized to the pp-wave background with two large angular momenta. The dependence on transverse coordinates is entirely dictated by the Tolman relation, whereas curvature corrections are suppressed by positive powers of $\beta$. 
To understand how these corrections are organized systematically, we next review the formal derivation of the thermal effective field theory on a curved background.

\paragraph{Thermal effective field theory and corrections.}
Following \cite{Banerjee:2012iz}, we consider a general stationary non-static background in $d\geq3$ spacetime dimensions.  After Wick rotation to Euclidean time, with $t_E\sim t_E+\beta$, the metric can be written in the standard  form
\begin{equation}
ds^2=N^2(\vec{x})(dt_E+a_i(\vec{x})dx^i)^2+h_{ij}dx^idx^j,
\end{equation}
where $a_i(\vec{x})$ is a Euclidean Kaluza–Klein connection, and the lapse $N(\vec{x})$ defines the local inverse temperature $\widetilde\beta=\beta N(\vec{x})$.

In the high-temperature regime,
the thermal wavelength is much smaller than the curvature scale of the background. Under the assumption that the dimensionally reduced thermal theory is gapped, all Matsubara modes can be integrated out. As a result, one obtains a $(d-1)$-dimensional hydrostatic generating functional $W=\log\mathcal{Z}$ depending
on static background fields and sources,
\begin{equation}
W[\widetilde\beta,h,a,\mathcal{J}]
=
\int d^{d-1}x\,\sqrt{h}\,\widetilde\beta\,
\mathcal{L}_{\mathrm{hydro}}(\widetilde\beta,h_{ij},a_i,\mathcal{J}).
\end{equation}

The one-point function is obtained by differentiating the generating functional with respect to the source,
\begin{equation}
\langle \mathcal O(\vec{x})\rangle_{\widetilde\beta}
=\frac{1}{\widetilde\beta\sqrt{h}}
\left.
\frac{\delta W}{\delta \mathcal{J}(\vec{x})}
\right|_{\mathcal{J}=0},
\end{equation}
so that its structure is entirely fixed by the derivative expansion of the generating functional.
For a conformal field theory, dimensional analysis strongly constrains the
possible terms. If the probe $\mathcal O$ has scaling dimension $\Delta_{\mathcal O}$,
the dimensionless source is $\widehat{\mathcal{J}} = \mathcal{J}\widetilde\beta^{d-\Delta_{\mathcal O}}$,
and the zero-derivative part of the hydrostatic effective action necessarily
takes the form
\begin{equation}
W^{(0)}
=
\int d^{d-1}x\,\sqrt{h}\,\widetilde\beta^{1-d}
P(\widehat{\mathcal{J}}),
\end{equation}
where $P$ is a dimensionless function. Expanding to first order in the source gives
\begin{equation}
P(\widehat{\mathcal{J}})=b_T+b_{\mathcal{O}}\widehat{\mathcal{J}}+O(\widehat{\mathcal{J}}^2),
\end{equation}
where $b_T$ and $b_{\mathcal O}$ are dimensionless coefficients depending only on the
underlying CFT, as described above.

This immediately gives
\begin{equation}
\langle \mathcal O(\vec{x})\rangle_{\widetilde\beta}^{(0)}
=
\frac{b_{\mathcal O}}{\widetilde\beta^{\Delta_{\mathcal O}}}.
\end{equation}

Similarly, one obtains the one-point function of the stress tensor,
\begin{equation}
\langle
T^{ij}(\vec{x})
\rangle_{\widetilde\beta}^{(0)}
=\frac{2}{\widetilde\beta\sqrt{h}}
\left.
\frac{\delta W^{(0)}}{\delta h_{ij}}
\right|_{\mathcal{J}=0}=b_T\widetilde\beta^{-d}h^{ij},
\end{equation}
and the energy density,
\begin{equation}
\varepsilon^{(0)}(\vec{x})
=-\frac{1}{\sqrt{h}}
\left.
\frac{\delta W^{(0)}}{\delta \widetilde\beta}
\right|_{\mathcal{J}=0}=(d-1)b_T\widetilde\beta^{-d}.
\end{equation}

The first corrections arise from the next terms in the derivative expansion.
Since the hydrostatic action is invariant under spatial diffeomorphisms and
parity, there are no one-derivative scalar invariants. The leading corrections contain two derivatives and are built from local geometric invariants
such as the spatial Ricci scalar and derivatives of the local inverse
temperature,
\begin{equation}
\langle \mathcal O\rangle_{\widetilde\beta}
=
\frac{b_{\mathcal O}}{\widetilde\beta^{\Delta_{\mathcal O}}}
+
\widetilde\beta^{2-\Delta_{\mathcal O}}
\left[
b_R R[h]
+
 b_1
\frac{\nabla_i\widetilde\beta\nabla^i\widetilde\beta}{\widetilde\beta^2}
+
b_2
\frac{\nabla^2\widetilde\beta}{\widetilde\beta}
+b_fN^2f_{ij}f^{ij}
\right]
+\cdots,
\label{eq:high-T-expansion}
\end{equation}
where $f_{ij}=\partial_ia_j-\partial_ja_i$.

This expression has a simple physical interpretation. The leading term
corresponds to the locally homogeneous thermal state, while the subleading
terms encode the response of the system to the slowly varying geometry. In
particular, the derivative expansion predicts that the first relative
corrections to the universal $\beta^{-\Delta_{\mathcal O}}$ scaling are
suppressed by $\beta^2$. Therefore, the corrections to the leading term in
equation ~\eqref{eq:highTexpected} are of relative order $\beta^2$.

\subsubsection{Low-temperature limit}

The low-temperature regime admits a complementary description governed by the lowest energy states. As  shown in \cite{Komargodski:2026ain}, this limit corresponds to the liberation regime on the cylinder, where states admit an approximate Fock-space description in terms of weakly interacting partons. At leading order, their twists are additive, and their Boltzmann weights factorize. Using this approximation, we compute the leading thermal one-point function of a scalar primary probe $\mathcal O$ from the CFT data.

First, we obtain the spectrum of $H_{\rm pp}$. As explained in Appendix~\ref{eq:Hinppwave}, a scalar primary $\mathcal P$ gives rise to a tower of pp-wave states labeled by the continuous null momentum $k>0$, together with transverse and longitudinal oscillator excitations. A basis of simultaneous eigenstates of $H_{\rm pp}$ and $P_x$ is denoted by $\lvert k,n_\alpha,s;\mathcal P\rangle$, with energies
\begin{equation}
E_{ n,s}(k)
=\frac{k}{2}+\tau_{\mathcal P}+n+2s,
\qquad n=\sum_{\alpha=1}^{d-2}n_\alpha.
\end{equation}

As $\beta\to\infty$, Boltzmann factors suppress the excited oscillator states, and the leading contribution from each conformal family comes from the $n=s=0$ trajectory, and the corresponding momentum eigenstates are given by the Fourier transform\footnote{See also \cite{Katz:2016hxp}. }
\begin{equation}
|k;\mathcal P\rangle= \frac{\mathcal{U}^{-1}}{\sqrt{\mathcal N_{\mathcal P}(k)}}
\int_{-\infty}^{\infty}
\frac{d\lambda}{2\pi}
\,e^{ik\lambda}
e^{-i\lambda P_x}
|\mathcal P\rangle,\qquad \langle k;\mathcal P|k';\mathcal P\rangle=\delta(k-k'),
\end{equation}
where $\mathcal N_{\mathcal P}(k)$ fixes the normalization, and $\mathcal{U}$ is a unitary transformation defined in \eqref{eq:app-single-twist} that commutes with $P_x$.

Consequently, at sufficiently low temperature, the leading contribution to the one-point function comes from the smallest-twist primary $\mathcal{P}$ that has a non-vanishing diagonal matrix element with $\mathcal O$, 
\begin{equation}
\langle\mathcal O(\vec y)\rangle_{\rm pp,\beta}
=
e^{-\beta\tau_{\mathcal P}}
\int_0^\infty
dk\,
e^{-\beta k/2}
\,
\langle k;\mathcal P|
\mathcal O(0,0,\vec y)
|k;\mathcal P\rangle
+\cdots,
\end{equation}
where the ellipsis denotes the contributions of heavier conformal families together with subleading thermal corrections.

Therefore, the problem is reduced to the computation of the diagonal matrix element
$\langle k;\mathcal P|
\mathcal O
|k;\mathcal P\rangle$,
which, remarkably, can be evaluated exactly using symmetries of the pp-wave and the fact that the unique vacuum is invariant under those (see Appendix \ref{ref:conformalinvariancePpwave}).
The normalization of the momentum eigenstates follows directly from normalization of energy eigenstates on the pp-wave  
\begin{equation}
\langle\mathcal P|
e^{i\lambda'P_x}
e^{-i\lambda P_x}
|\mathcal P\rangle
=
\frac{
C_{\mathcal P}
}{
\left[
(-i)(\lambda'-\lambda+i0)
\right]^{\tau_{\mathcal P}}
}.
\end{equation}
Performing the Fourier transform immediately yields
\begin{equation}
\mathcal N_{\mathcal P}(k)
=
C_{\mathcal P}
\frac{
k^{\tau_{\mathcal P}-1}
}{
\Gamma(\tau_{\mathcal P})
}.
\label{2pt-gen}
\end{equation}

Substituting the definition of the momentum eigenstates, the diagonal matrix element becomes
\begin{equation}
\langle k;\mathcal P|
\mathcal O(\vec y)
|k;\mathcal P\rangle
=
\frac{1}{\mathcal N_{\mathcal P}(k)}
\int
\frac{d\lambda}{2\pi}
\frac{d\lambda'}{2\pi}
\,e^{ik(\lambda-\lambda')}
\,
\mathcal G_{\mathcal P\mathcal P\mathcal O}
(\lambda,\lambda';\vec y).
\end{equation}

Conformal invariance on the pp-wave (in the sense of Appendix \ref{ref:conformalinvariancePpwave}) fixes this three-point function uniquely up to the OPE coefficient,
\begin{equation}
\mathcal G_{\mathcal P\mathcal P\mathcal O}
(\lambda,\lambda';\vec y)
=
\frac{C_{\mathcal P\mathcal P\mathcal O}
e^{-i\pi(\tau_{\mathcal P}-\Delta_{\mathcal O}/2)/2}
}{
(\lambda-i|\vec y|^2/2)^{\Delta_{\mathcal O}/2}
(\lambda'+i|\vec y|^2/2)^{\Delta_{\mathcal O}/2}
(\lambda-\lambda'-i0)^{\tau_{\mathcal P}-\Delta_{\mathcal O}/2}
}.
\end{equation}

 The resulting one-particle kernel is
\begin{multline}
\label{eq:one-part-kernel}
\langle k;\mathcal P|
\mathcal O(\vec y)
|k;\mathcal P\rangle
=
\frac{
C_{\mathcal P\mathcal P\mathcal O}
}{
C_{\mathcal P}
}
\frac{
\Gamma(\tau_{\mathcal P})
\Gamma(\Delta_{\mathcal O}-1)
}{
\Gamma(\Delta_{\mathcal O}/2)^2
\Gamma(\tau_{\mathcal P}+\Delta_{\mathcal O}/2-1)
}
\\
\times
k^{\Delta_{\mathcal O}/2-1}
\,{}_1F_1
\!\left(
\Delta_{\mathcal O}-1;
\tau_{\mathcal P}+\frac{\Delta_{\mathcal O}}{2}-1;
-k|\vec y|^2
\right),\qquad \Delta_{\mathcal O}>1.
\end{multline}
Inserting this expression into the thermal trace and performing the remaining integral over the momentum $k$, we obtain the leading low-temperature contribution to the one-point function,
\begin{empheq}{multline}
\langle
\mathcal O(\vec y)
\rangle_{\rm pp,\beta}
\overset{\beta\to\infty}{\simeq}
e^{-\beta\tau_{\mathcal P}}
\left(\frac{2}{\beta}\right)^{\Delta_{\mathcal O}/2}
\frac{
C_{\mathcal P\mathcal P\mathcal O}
}{
C_{\mathcal P}
}
\frac{
\Gamma(\tau_{\mathcal P})
\Gamma(\Delta_{\mathcal O}-1)
}{
\Gamma(\Delta_{\mathcal O}/2)
\Gamma(\tau_{\mathcal P}+\Delta_{\mathcal O}/2-1)
}
\\
\times
{}_2F_1
\!\left(
\Delta_{\mathcal O}-1,
\frac{\Delta_{\mathcal O}}{2};
\tau_{\mathcal P}+\frac{\Delta_{\mathcal O}}{2}-1;
-\frac{2|\vec y|^2}{\beta}
\right),\qquad \Delta_{\mathcal O}>1.
\label{eq:1f_1pt_pp}
\end{empheq}

Equation~\eqref{eq:1f_1pt_pp} provides a universal prediction for the leading low-temperature behavior of one-point functions on the pp-wave background. Once the conformal data of the lightest contributing conformal family,
\(
(\tau_{\mathcal P},\,C_{\mathcal P},\,C_{\mathcal P\mathcal P\mathcal O}),
\)
are specified, the leading thermal correction is completely determined. The  Boltzmann factor reflects the energy gap to the lightest state contributing to the thermal trace, while the dependence on the transverse position is encoded in the hypergeometric function.

Finally, we stress an important limiting case of this result when the scaling dimension of the probe satisfies $\Delta_{\mathcal{O}}=2\tau_{\mathcal{P}}$.
In this case, the one-particle kernel collapses to 
\begin{equation}
 \langle k;\mathcal P|\mathcal O(\vec y)|k;\mathcal P\rangle
=\frac{
C_{\mathcal P\mathcal P\mathcal O}
}{
C_{\mathcal P}
}\frac{
k^{\tau_{\mathcal P}-1}
}{
\Gamma(\tau_{\mathcal P})
}e^{-k|\vec{y}|^2},
\end{equation}
and the thermal one-point function is described by a simple power law,
\begin{equation}
\label{eq:special-one-point}
\langle
\mathcal O(\vec y)
\rangle_{\rm pp,\beta}
\overset{\beta\to\infty}{\simeq}\frac{
C_{\mathcal P\mathcal P\mathcal O}
}{
C_{\mathcal P}}\frac{e^{-\beta\tau_{\mathcal P}}}{(\beta/2+|\vec{y}|^2)^{\tau_{\mathcal P}}},\qquad \Delta_{\mathcal{O}}=2\tau_{\mathcal{P}}\geq 1.
\end{equation}

\subsection{Two large angular momenta in (3+1)d}\label{sec:two-large-spins}

The discussion extends straightforwardly to the case of two large angular momenta. For a four-dimensional CFT, we define
$\tau=\Delta-J_1-J_2$. The corresponding pp-wave geometry is
\begin{equation}
    ds_{\rm pp_2}^2
=-dt^2-2 dt dv-4 u dt dy +du^2+dy^2.
\end{equation}

Compared to the one-large-spin background, the symmetry algebra is considerably enhanced. Besides the Hamiltonian generating translations in $t$, the isometry group of the pp-wave above is the Heisenberg group $\mathbb H_5$. However at finite temperature only a subset of those survive: this consists of time translations, $\partial_t$, and $\mathbb H_3 \subset \mathbb H_5$ 
with generators \begin{equation}
    K_{\pm} = \frac{1}{\sqrt 2}\left[\partial_y\pm  i (\partial_u-2 y \partial_v)\right]
\end{equation} and $\partial_v$ satisfying
\begin{equation}
[K_+,K_-]
=
2i\,\partial_v.
\end{equation}

 In addition, the generators $K_\pm$ transform as a doublet under an $SO(2)$ automorphism corresponding to rotations in the transverse plane.

The crucial observation is that both the Heisenberg generators and the $SO(2)$ rotations commute with the pp-wave Hamiltonian. Consequently, unlike in the one-large-spin background, the $\mathbb H_3$ Heisenberg subgroup remains unbroken at finite temperature. Since scalar operators are singlets under these symmetries, their thermal expectation values cannot depend on the transverse coordinates. Therefore, the one-point functions are necessarily constant throughout the pp-wave geometry,
a fact that considerably simplifies the structure of the high-temperature expansion in the two-spin case.
Matching this prediction with the high temperature prediction, one concludes that \begin{equation}
\label{eq:high-T-two-spins}
    \langle\mathcal O\rangle_{\rm pp_2,\beta}\overset{\beta\to 0}{\simeq}\frac{b_{\mathcal O}}{\beta^{\Delta_{\mathcal O}}}\left[1+O(\beta^2)\right],
\end{equation}
where $b_{\mathcal O}$ is the thermal one-point function's coefficient at infinite temperature.

In the low-temperature limit, the leading contribution comes from the minimal twist primary $\mathcal{P}$. The spatial symmetries discussed above make the thermal one-point function independent
of position, so we evaluate the probe at the origin. As shown in the Appendix, the transverse excitations organize into Landau levels, with the energy spectrum\footnote{The existence of the twist gap in the $2+1$ dimensional case, or more generally with only one spin considered large, is ensured and predicted by analytic bootstrap arguments. This is not the case for the $3+1$ dimensional case if we consider both $J_1$ and $J_2$ large. Nonetheless we expect this property in strongly coupled theories and we therefore assume the existence of the twist gap throughout this Section. We thank Jeremy Mann for discussions on this point.}
\begin{equation}
    E_{n,s}(k)=\frac{k}{2}+2(n+s)+\tau_{\mathcal{P}}.
\end{equation}
For each fixed $s$, the integer $n$ labels the Landau level, while
$\kappa=0,1,2,...$ labels its degenerate states. The integer $s$ labels an additional tower of excitations within the scalar conformal family. At low temperature, the leading contribution comes
from the lowest Landau level in the lowest such tower, $n=s=0$. The leading contribution to the thermal one-point function becomes
\begin{equation}
  \langle\mathcal O\rangle_{\rm pp_2,\beta}
\overset{\beta\to\infty}{\simeq}
e^{-\beta\tau_{\mathcal P}}
\int_0^\infty
dk\,
e^{-\beta k/2}
\sum_{\kappa=0}^\infty\,
\langle k,\kappa;\mathcal P|
\mathcal O(0)
|k,\kappa;\mathcal P\rangle,
\end{equation}
where the state $|k,\kappa\rangle$ is associated with the lowest Landau level. As in the one spin case, the leading order kernel is determined by the flat-space three-point function $\langle\mathcal{P}^\dagger\mathcal{O}\mathcal{P}\rangle$, with the normalization fixed by $\langle\mathcal{P}^\dagger\mathcal{P}\rangle$. Explicitly, we find in Appendix \ref{sec:LLL-kernel},
\begin{equation}
\label{eq:app-2-spin-kernel}
\langle k,\kappa;\mathcal P|
\mathcal O
|k,\kappa;\mathcal P\rangle=\frac{(\tau_{\mathcal P}-\Delta_{\mathcal O}/2)_\kappa}{(\tau_{\mathcal P}+\Delta_{\mathcal O}/2-1)_\kappa}\langle k,0;\mathcal P|
\mathcal O
|k,0;\mathcal P\rangle,
\end{equation}
and
\begin{equation}
\langle k,0;\mathcal P|
\mathcal O
|k,0;\mathcal P\rangle=\frac{C_{\mathcal P\mathcal P\mathcal O}}{C_{\mathcal P}}\frac{
        \Gamma(\tau_{\mathcal P})
        \Gamma(\Delta_{\mathcal O}-1)
    }{
        \Gamma(\Delta_{\mathcal O}/2)^2
        \Gamma(\tau_{\mathcal P}+\Delta_{\mathcal O}/2-1)}k^{\Delta_{\mathcal O}/2-1},
\end{equation}
where $(b)_\kappa$ is the Pochhammer symbol.
Now, it remains to sum over degenerate states and integrate over momentum $k$. At large $\kappa$, the kernel scales as $O(\kappa^{1-\Delta_{\mathcal O}})$, so the sum is absolutely convergent for $\Delta_{\mathcal O}>2$,
\begin{equation}
\sum_{\kappa=0}^\infty\frac{(\tau_{\mathcal P}-\Delta_{\mathcal O}/2)_\kappa}{(\tau_{\mathcal P}+\Delta_{\mathcal O}/2-1)_\kappa}=\frac{\tau_{\mathcal P}+\Delta_{\mathcal O}/2-2}{\Delta_{\mathcal O}-2},
\end{equation}
and
\begin{equation}
\int_0^\infty dk  k^{\Delta_{\mathcal O}/2-1}e^{-\beta k/2}=\Gamma(\Delta_{\mathcal O}/2)\Big(\frac{2}{\beta}\Big)^{\Delta_{\mathcal O}/2}.
\end{equation}
Using these formulas, we obtain the thermal one-point function on the two-spin pp-wave background,
\begin{equation}
\label{eq:low-T-two-spins}
\langle\mathcal O\rangle_{{\rm pp_2},\beta}
\overset{\beta\to\infty}{\simeq}
e^{-\beta\tau_{\mathcal P}}
\left(\frac{2}{\beta}\right)^{\Delta_{\mathcal O}/2}
\frac{C_{\mathcal P\mathcal P\mathcal O}}{C_{\mathcal P}}
\frac{
\Gamma(\tau_{\mathcal P})
\Gamma(\Delta_{\mathcal O}-2)
}{
\Gamma(\Delta_{\mathcal O}/2)
\Gamma(\tau_{\mathcal P}+\Delta_{\mathcal O}/2-2)
},\qquad \Delta_{\mathcal O}>2.
\end{equation}
In the special case $\Delta_{\mathcal O}=2\tau_{\mathcal P}$, this takes the form
\begin{equation}\label{eq:low-T-special}
\langle\mathcal O\rangle_{{\rm pp_2},\beta}
\overset{\beta\to\infty}{\simeq}
e^{-\beta\tau_{\mathcal P}}
\left(\frac{2}{\beta}\right)^{\tau_{\mathcal P}}\frac{C_{\mathcal P\mathcal P\mathcal O}}{C_{\mathcal P}}
,\qquad \Delta_{\mathcal O}=2\tau_{\mathcal P}\geq2.
\end{equation}

In section \ref{sec:weyl-laplace}, we show how the above results can be Laplace transformed to obtain the asymptotic three point function given in \eqref{eq:asymptotics3+1}.

\section{Examples}\label{sec:examples}
In this section, we study particular examples of CFTs on the pp-wave geometry, provide an independent computation of thermal one-point functions by solving equations of motion, and compare the results with our general ansatz.

\subsection{Free scalar on pp-wave}
We start with showing how the prediction produced above matches the simplest possible theory, i.e. the example of the free scalar theory.

\paragraph{$\langle {:}\phi^2{:}\rangle_{\rm pp,\beta}$ in 2+1 dimensions.}

A free massless scalar on the $(2+1)d$ pp-wave background satisfies the equation of motion
\begin{equation}
    \left[2\partial_t\partial_x+(1+y^2)\partial_x^2+\partial_y^2\right]\phi=0.
\end{equation}
For a positive-frequency mode $e^{-iEt+ikx}\varphi(y)$ with $k>0$, this becomes
\begin{equation}
 \left[-\partial_y^2+k^2y^2\right]\varphi
 =(2kE-k^2)\varphi.
\end{equation}
Therefore, the transverse motion is described by harmonic oscillations, with energies and normalized wavefunctions
\begin{equation}
 E_{n}(k)=\frac{k}{2}+n+\frac12 ,
 \qquad
 \varphi_{k,n}(y)=
 \frac{k^{1/4}}{\pi^{1/4}\sqrt{2^n n!}}
 H_n\big(\sqrt{k}y\big)e^{-ky^2/2} .
 \label{eq:Psi}
\end{equation}
Accordingly, the canonically normalized field expansion is
\begin{equation}
 \phi(t,x,y)=\int_0^\infty\frac{dk}{2\pi}\sum_{n=0}^\infty
 \frac{\varphi_{k,n}(y)}{\sqrt{2k}}
 \left[b_{k,n}e^{-iE_n(k)t+ikx}
 +b_{k,n}^{\dagger}e^{iE_n(k)t-ikx}\right],
\end{equation}
where $[b_{k,n},b_{k',n'}^\dagger]=2\pi\delta(k-k')\delta_{nn'}$.

We consider a one-point function of the following scalar primary operator $\mathcal{O}={:}\phi^2{:}$, where normal ordering is defined relative to the zero-temperature pp-wave vacuum. Evaluating the thermal trace gives
\begin{equation}
\langle {:}\phi^2{:}(y)\rangle^{}_{\rm pp,\beta} = \int_0^\infty \frac{dk}{2 \pi k}\sum_{n = 0}^\infty \frac{|\varphi_{k,n}(y)|^2}{e^{\beta (k/2+n+1/2)}-1}.
\end{equation}
Expanding the Bose-Einstein distribution and using Mehler's resummation gives the closed form expression
\begin{equation}
\langle {:}\phi^2{:}(y)\rangle^{}_{\rm pp,\beta}
=
\frac{1}{2\pi}
\sum_{m=1}^{\infty}
\frac{1}{\sqrt{\sinh(m\beta)
\big[m\beta+2y^2\tanh({m\beta}/{2})\big]}}
 .
\label{eq:pp-wave-1pt-free-3d}
\end{equation}
At low temperature, the sum is dominated by the lowest mode, $m=1$,
\begin{equation}
 \langle {:}\phi^2{:}(y)\rangle^{}_{\rm pp,\beta} \overset{\beta\to\infty}{=} \frac{e^{-\beta /2}}{2\pi\sqrt{\beta/2+y^2}},
 \label{eq:free-3d-lowT}
\end{equation}
where the suppressing Boltzmann factor is set by the scalar twist $\tau_\phi=1/2$. This result is in perfect agreement with the general prediction \eqref{eq:special-one-point}.

The high-temperature limit is slightly subtler because the sum over winding sectors does not commute with the naive $\beta\to0$ expansion.
In fact, naively we would have 
\begin{equation}
    \langle {:}\phi^2{:}(y)\rangle_{\rm pp,\beta} \overset{\rm naive}{=} \frac{1}{2 \pi}\sum_{m = 1}^\infty \frac{1}{m \beta \sqrt{1+y^2}},
\end{equation}
which diverges. The trick is to rewrite it as \begin{multline}
    \langle {:}\phi^2{:}(y)\rangle_{\rm pp,\beta} = \frac{1}{2 \pi}\sum_{m = 1}^\infty \frac{e^{-m \beta }}{m \beta \sqrt{1+y^2}}+\\+\frac{1}{2 \pi}\sum_{m = 1}^\infty\left[ \frac{1}{\sqrt{\sinh(m\beta)
\big[m\beta+2y^2\tanh({m\beta}/{2})\big]}}-\frac{e^{-m \beta }}{m \beta \sqrt{1+y^2}}\right].
\end{multline}
The first term, capturing the $m \beta$ singularity,  is now convergent while the second is a Riemann sum when $\beta \to 0$.
Therefore we obtain 
\begin{equation}
\langle {:}\phi^2{:}(y)\rangle^{}_{\rm pp,\beta}\overset{\beta\to0}{=}\frac{1}{\beta}\left[\frac{\log \left(1/\beta\right)}{2 \pi  \sqrt{1+y^2}}+ { \mathcal I(y)}\right],
\end{equation}
where the logarithm reflects the infrared sensitivity of the massless scalar in two spatial dimensions, and $\mathcal I(y)$ is a convergent integral \begin{equation}
    \mathcal I(y) = \frac{1}{2 \pi}\int_0^\infty dx \ \left[\frac{1}{\sqrt{\sinh x \left[x+2 y^2 \tanh(x/2)\right]}}- \frac{e^{-x}}{x \sqrt{1+y^2}}\right].
\end{equation}

\paragraph{$\langle {:}\phi^2{:}\rangle_{\rm pp_2,\beta}$ in 3+1 dimensions.}

The free scalar on the $(3+1)d$ two large spin pp-wave background provides a complementary example in which the transverse dynamics is a Landau-level problem rather than a single harmonic oscillator. Explicitly,
a positive-frequency mode $e^{-iEt-ikv+i\kappa y}\varphi(u)$ satisfies
\begin{equation}
 \left[-\partial_u^2+(2ku+\kappa)^2\right]\varphi
 =(2kE-k^2)\varphi,
 \qquad E_n(k)=\frac{k}{2}+2n+1.
\end{equation}
The energy spectrum is independent of the momentum $\kappa$, which labels the center of oscillations.
Integrating over the centers gives the Landau
degeneracy
\begin{equation}
 \int_{-\infty}^{\infty}\frac{d\kappa}{2\pi}
 \left|\varphi_{k,n}\!\left(u+\frac{\kappa}{2k}\right)\right|^2
 =\frac{k}{\pi},
\end{equation}
and removes the position dependence of the thermal one-point function, 
\begin{align}
 \langle{:}\phi^2{:}\rangle_{\mathrm{pp}_2,\beta}
 &=\int_0^\infty\frac{dk}{2\pi^2}
\sum_{n=0}^\infty\frac1{e^{\beta(k/2+2n+1)}-1}\notag\\
&=\frac1{2\pi^2\beta}\sum_{m=1}^\infty\frac1{m\sinh(m\beta)}
 =-\frac1{\pi^2\beta}\sum_{n=0}^\infty
 \log\!\left(1-e^{-(2n+1)\beta}\right).
\end{align}
The low-temperature limit is dominated by the lowest Landau level, $n=0$, whereas the high-temperature limit reproduces the flat-space expectation values for one real scalar with the coefficient $b_{\phi^2}=1/12$,
\begin{equation}
 \langle{:}\phi^2{:}\rangle_{\mathrm{pp}_2,\beta}
\simeq
 \begin{cases}
 \displaystyle\frac{e^{-\beta}}{\pi^2\beta},&\beta\gg1,\\[11pt]
 \displaystyle\frac1{12\beta^2}-\frac{\log{2}}{2\pi^2\beta}+\frac{1}{24\pi^2},&\beta\ll1.
 \end{cases}
\end{equation}
The leading-order results are in agreement with our general predictions \eqref{eq:low-T-special} and \eqref{eq:high-T-two-spins}.
However, at high temperature, the first subleading term $-{\log{2}}/({2\pi^2\beta})$ is of relative order $O(\beta)$ rather than $O(\beta^2)$, which has a simple explanation. 
Indeed, the EFT description in \ref{sec:high-T} assumes that thermal correlations must be sufficiently short-ranged to have a local effective action involving only background fields. The free massless scalar does not satisfy this assumption, since it retains an unscreened Matsubara zero mode (see also \cite{Benjamin:2023qsc}).

\subsection{\texorpdfstring{$\mathrm{O}(N)$ model at large $N$}{O(N) model at large N}}

As a first non-trivial application, we consider the critical $\mathrm{O}(N)$ model in the large-$N$ limit. This theory provides an ideal testing ground for our general formulas, since the thermal one-point functions can be computed exactly by solving the large-$N$ saddle. In particular, it allows us to test both the high-temperature expansion \eqref{eq:highTexpected} and the low-temperature formula \eqref{eq:1f_1pt_pp}.

The critical $\mathrm{O}(N)$ model can be conveniently described by introducing an auxiliary Hubbard--Stratonovich field $\sigma$, whose Euclidean Lagrangian is
\begin{equation}
\mathcal L_E
=
\frac12(\partial\phi_i)^2
+
\frac{1}{2}\sigma\,\phi_i\phi_i,
\end{equation}
where the index $i=1,\ldots,N$ is summed over. At leading order in the large-$N$ expansion, the dynamics is completely determined by the saddle-point value of the auxiliary field. In particular, the thermal one-point function of the Hubbard--Stratonovich field is defined by
\begin{equation}
\langle\sigma(y)\rangle_{\rm pp,\beta}\simeq \sigma_*(y,\beta)+O(N^{-1}).
\end{equation}
In flat space at zero temperature, conformal invariance implies that the saddle is simply
\begin{equation}
\sigma_*=0.
\end{equation}
However, at finite temperature, the thermal background generates a non-trivial expectation value for $\sigma_*$ (see \cite{Sachdev:1992py,Petkou:2018ynm}). On the pp-wave geometry, the saddle generally depends both on the inverse temperature $\beta$ and on the transverse coordinate $y$. To compute this profile, we solve the large-$N$ gap equation,
\begin{equation}
G_{\sigma_*}^{\beta}(y;y)
-
G_{0}^{\infty}(y;y)
=
0,
\label{eq:gap}
\end{equation}
where $G_{\sigma_*}^{\beta}$ denotes the thermal propagator of the fundamental field evaluated in the background $\sigma_*(y)$, while $G_{0}^{\infty}$ is the zero-temperature propagator in flat space. The subtraction removes the ultraviolet divergence and implements the criticality condition.

From the CFT point of view, Eq.~\eqref{eq:gap} admits a simple interpretation. In the interacting critical theory, the composite operator $\phi_i\phi_i$ is replaced by the auxiliary field $\sigma$ through the Hubbard--Stratonovich construction. Consequently, the gap equation is equivalent to imposing the absence of the operator $\phi_i\phi_i$ in the short-distance expansion of two fundamental fields, ensuring that the theory remains at the conformal fixed point even in the presence of the thermal background.

\begin{figure}[t!]
  \centering
  \begin{subfigure}[t]{0.495\textwidth}
    \centering
    \includegraphics[width=\linewidth]{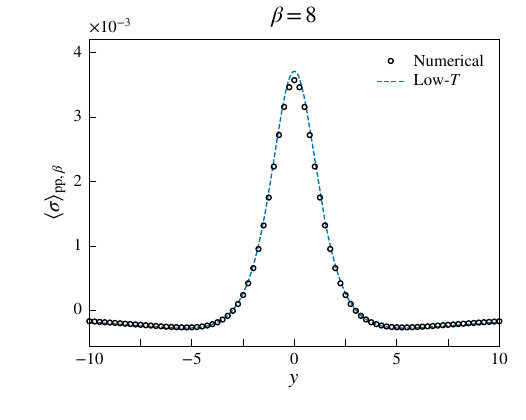}
    \caption{}
  \end{subfigure}
  \hfill
  \begin{subfigure}[t]{0.495\textwidth}
    \centering
\includegraphics[width=\linewidth]{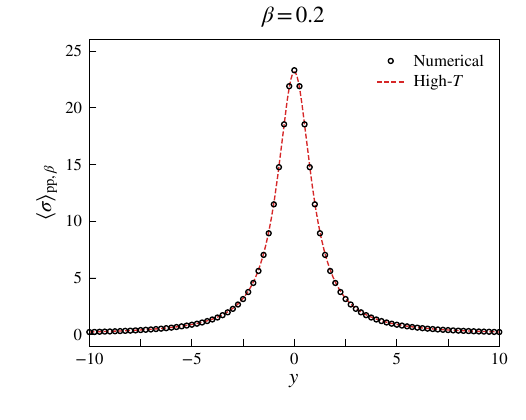}
    \caption{}
  \end{subfigure}
 \caption{Profile of the Hubbard–Stratonovich field in the low-temperature (a) and high-temperature (b) regimes. The analytical predictions (dashed lines) are compared with the numerical results.}
 \label{Fig:sigmaprofile}
\end{figure}

The gap equation \eqref{eq:gap} does not admit a closed-form solution at finite temperature and must be solved numerically. Nevertheless, both the high- and low-temperature limits can be analyzed analytically and provide non-trivial checks of the general formulas derived in the previous sections.

\paragraph{High-temperature limit.}

At sufficiently high temperatures, the thermal wavelength is much shorter than the characteristic curvature scale of the pp-wave background. The solution is therefore locally approximated by the homogeneous thermal saddle in flat space, whose expectation value is (see \cite{Sachdev:1992py,Petkou:2018ynm})
\begin{equation}
\langle\sigma\rangle_{\beta}
\simeq
\frac{4\log^2\!\left(\frac{1+\sqrt5}{2}\right)}
{\beta^2}.
\end{equation}
The only effect of the curved background is the local redshift of the temperature discussed in the previous section in Eq. \eqref{eq:high-T-one-point}, leading to
\begin{equation}
\langle\sigma(y)\rangle_{\rm pp,\beta}
\simeq
\frac{
4\log^2\!\left(\frac{1+\sqrt5}{2}\right)
}{
\beta^2(1+y^2)
},
\end{equation}
which is in total agreement with the general high-temperature prediction \eqref{eq:highTexpected}.

\paragraph{Low-temperature limit.}

In the opposite regime, the saddle approaches the conformal vacuum and the auxiliary field becomes exponentially small. It is therefore sufficient to linearize the gap equation around
\(
\sigma_*=0,
\)
obtaining
\begin{equation}
0
=
G^\beta_{\sigma_*}(y;y)
-
G^\infty_0(y;y)
\simeq
G^{\rm th}_0(y;y)
-
\int dy'\,
\Pi(y;y')
\sigma_*(y',\beta),
\end{equation}
where
\begin{equation}
G^{\rm th}_0(y;y)
=
G^\beta_0(y;y)
-
G^\infty_0(y;y)
\end{equation}
is the thermal part of the free propagator. Using the general pp-wave thermal propagator derived previously, one finds
\begin{multline}
G^{\rm th}_0(y;y)
=
\frac1N
\sum_i
\langle{:}\phi_i^2{:}\rangle_\beta
\\
=
\sum_n
\int_0^\infty
\frac{dk}{2\pi k}
\frac{
|\varphi_{k,n}(y)|^2
}{
e^{\beta(k/2+n+1/2)}-1
}
=
\frac{
e^{-\beta/2}
}{
2\pi\sqrt{\beta/2+y^2}
}
+
O(e^{-\beta}),
\end{multline}
while
\begin{equation}
\Pi(y,y')
=
\int
dt'\,
dx'\,
G^\infty_0(y;t',x',y')
G^\infty_0(t',x',y';y)
\end{equation}
is the standard large-$N$ polarization bubble.
Passing to momentum space diagonalizes the integral equation,
\begin{equation}
\widetilde G^{\rm th}_0(q)
=
\widetilde\Pi(q)
\widetilde\sigma_*(q,\beta),
\end{equation}
and using the critical bubble
\begin{equation}
\widetilde\Pi(q)
=
\frac{1}{8|q|}
\end{equation}
gives
\begin{equation}
\widetilde\sigma_*(q,\beta)
=
\frac{8e^{-\beta/2}}{\pi}
|q|
K_0
\!\left(
|q|
\sqrt{\frac\beta2}
\right)
+
O(e^{-\beta}).
\end{equation}
The inverse Fourier transform then yields\footnote{See also Eqs. (4.56)–(4.57) in \cite{mishra2026semiuniversalityonmodels1times}.}
\begin{equation}
\sigma_*(y,\beta)
=
\frac{8e^{-\beta/2}}{\pi^2}
\left[
\frac1{\beta/2+y^2}
-
\frac{
y\,\operatorname{arcsinh}(y\sqrt{2/\beta})
}{
(\beta/2+y^2)^{3/2}
}
\right]
+
O(e^{-\beta}).
\end{equation}

This result provides a highly non-trivial check of the general low-temperature formula. Indeed, substituting
\(
\Delta_\sigma=2
\)
and
\(
\tau_\phi=1/2
\)
into Eq.~\eqref{eq:1f_1pt_pp} immediately gives
\begin{equation}
\langle\sigma(y)\rangle_{\rm pp,\beta}
\overset{\beta\to\infty}{=}
e^{-\beta/2}
\frac{2}{\beta}
    \sum_i\frac{C_{\phi_i\phi_i\sigma}}{C_{\phi_i}}
\,{}_2F_1
\!\left(
1,1;
\frac12;
-\frac{2y^2}{\beta}
\right),
\end{equation}
which exactly reproduces the solution of the linearized gap equation. We also compute the profile numerically and show the results in Fig.~\ref{Fig:sigmaprofile}.

\paragraph{Free energy density.}
Once the saddle-point profile $\sigma_*(y)$ has been determined, the free energy follows immediately. At leading order in the large-$N$ expansion, the dynamics of the fundamental fields reduces to a one-particle Schrödinger problem. For each value of the conserved null momentum $k$, the relevant Hamiltonian is
\begin{equation}
H_k[\sigma_*]
=
-\partial_y^2
+
k^2y^2
+
\sigma_*(y),
\end{equation}
whose eigenvalues $\lambda_n(k,\sigma_*)$ determine the single-particle energies,
\begin{equation}
E_n(k,\beta)
=
\frac{k}{2}
+
\frac{\lambda_n(k,\sigma_*)}{2k}.
\end{equation}
 The free energy density is then obtained by summing the Bose-Einstein contributions of all normal modes and subtracting the renormalized vacuum energy density,
\begin{equation}
\frac{\log Z}{\beta NV_x}=\mathcal{F}(\beta)=
-\frac{1}{\beta}\int_0^\infty
\frac{dk}{2\pi}
\sum_n
\log
\!\left(
1-e^{-\beta E_n(k,\beta)}
\right)- \mathcal{E}^{\rm ren}_{\rm vac}[\sigma_*].
\label{eq:ON-free-energy}
\end{equation}
Using the saddle-point condition, the thermal and renormalized vacuum contributions can be combined into a finite spectral expression,
\begin{equation}
\mathcal{F}(\beta)=\frac{1}{2}\int_0^\infty \frac{dk}{2\pi}\sum_n\frac{k}{e^{\beta E_n(k,\beta)}-1}.
\end{equation}

\begin{figure}[t]
    \centering
    \includegraphics[scale=1.05]{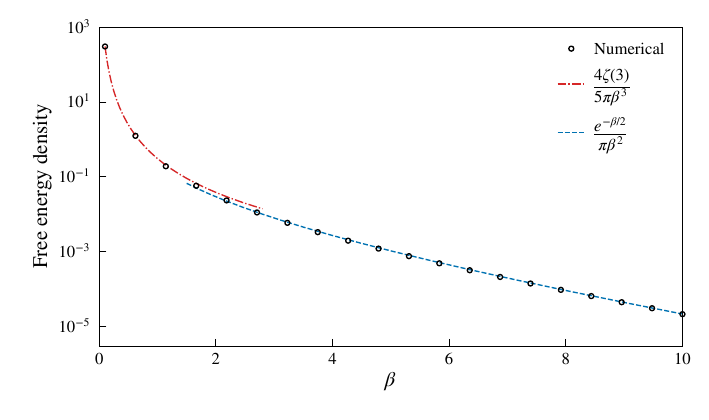}
    \caption{Free energy density of the large-$N$ $\mathrm O(N)$ model. The numerical solution obtained from the saddle-point profile is compared with the analytic high- (dash-dotted red line) and low-temperature  (dashed blue line) asymptotics.}
    \label{fig:FONmodel}
\end{figure}

The asymptotic behavior of the free energy follows directly from the general arguments discussed in \cite{Komargodski:2026ain}. At high temperature, the leading
contribution is obtained from the flat-space pressure \cite{Sachdev:1992py,Sachdev:1993polylog,Barrat:2025wbi,David:2025tqn}, evaluated at the local inverse temperature $\widetilde\beta(y)$. Integrating over the transverse coordinate gives
\begin{equation}
\mathcal{F}(\beta)\overset{\beta\to0}{\simeq}
\frac{2\zeta(3)}{5\pi\beta^3}
\int_{-\infty}^{\infty}\frac{dy}{(1+y^2)^{3/2}}
=\frac{4\zeta(3)}{5\pi\beta^3}.
\end{equation}
In the opposite limit, the saddle approaches the vacuum solution
$\sigma_*=0$, and the leading contribution comes from the lowest
excitation band, with $E_0(k)=k/2+1/2$, giving
\begin{equation}
\mathcal{F}(\beta)
\overset{\beta\to\infty}{\simeq}\frac{e^{-\beta/2}}{\pi\beta^2}.
\end{equation}

In Figure~\ref{fig:FONmodel}, we compare these analytic predictions with the numerical evaluation of Eq.~\eqref{eq:ON-free-energy} and find excellent agreement in both asymptotic regimes. The numerical solution smoothly interpolates between them. In particular, the numerical free energy density appears smooth throughout the sampled crossover, providing no indication of a phase transition on the pp-wave background.

\subsection{Free fermions on pp-wave}
\label{sec:pp-free-fermion}

So far, we have discussed scalar probe operators that are even under parity transformation. Let us now study the thermal one-point functions of a pseudoscalar ${\mathcal O}$ on the $(2+1)d$ pp-wave background.
Since the reflection $y\mapsto-y$ is an isometry of the pp-wave background and preserves the thermal state, its one-point function must satisfy
\begin{equation}
\langle{\mathcal O}(y)\rangle_{\rm pp,\beta}=-\langle{\mathcal O}(-y)\rangle_{\rm pp,\beta},
\end{equation}
allowing a nonzero odd profile that vanishes at $y=0$.

As the simplest example, we consider free massless Dirac fermions $\psi$ and take ${\mathcal O}={:}\bar{\psi}\psi{:}$, where normal ordering is defined with respect to the zero-temperature pp-wave vacuum. We use
\begin{equation}
\gamma^0=i\sigma_2,\qquad \gamma^1=\sigma_1 ,\qquad \gamma^2=\eta\sigma_3,\qquad \bar{\psi}=i\psi^\dagger\gamma^0,\qquad \eta=\pm1,
\end{equation}
where $\sigma_i$ are the Pauli matrices, and the identity $\gamma^0\gamma^1\gamma^2=\eta\mathbf{1}$ distinguishes the two irreducible Clifford representations.
With these spinor conventions, the massless Dirac equation on the pp-wave becomes
\begin{equation}
 \begin{pmatrix}
  \eta\partial_y &
  \sqrt2\bigl(\partial_t+\tfrac12(1+y^2)\partial_x\bigr)\\
  \sqrt2\partial_x & -\eta\partial_y
 \end{pmatrix}\psi=0.
\end{equation}
Expanding the Dirac fermion into positive and negative-frequency modes, $e^{-iEt+ikx}u_{k,n}$ and $e^{iEt-ikx}v_{k,n}$, we find the same spectrum as in the free scalar case: $E_n(k)={k}/{2}+n+1/2$.

Now we proceed as in the free scalar calculation. Expanding the Fermi-Dirac distribution and using Mehler’s resummation, we obtain
\begin{equation}
 \langle {:}\bar{\psi}\psi{:}(y)\rangle_{\mathrm{pp},\beta}
 =\frac{\eta y}{4\pi}\sum_{m=1}^\infty \frac{(-1)^{m+1}}
 {\sinh(m\beta/2)\big[y^2+
   \frac{m\beta}{2}\coth(m\beta/2)\big]^{3/2}}.
 \label{eq:ferm-sum}
\end{equation}
At low temperature, the lowest winding mode $m=1$ dominates, and we immediately get
\begin{equation}
\langle {:}\bar{\psi}\psi{:}(y)\rangle_{\mathrm{pp},\beta}
\overset{\beta\to \infty}{\simeq}
\frac{\eta e^{-\beta/2}y}{2\pi(\beta/2+y^2)^{3/2}}.
\end{equation}
And at high temperature, we analogously obtain
\begin{equation}
\langle {:}\bar{\psi}\psi{:}(y)\rangle_{\mathrm{pp},\beta}\overset{\beta\to 0}{\simeq}
\frac{\eta\log{2}}{2\pi\beta}\frac{y}{(1+y^2)^{3/2}}.
\label{eq:high-odd}
\end{equation}

\paragraph{High-temperature EFT for pseudoscalars.}

The temperature dependence in \eqref{eq:high-odd} has a simple interpretation, complementing the parity-even scalar expansion from section \ref{sec:high-T}. In a homogeneous flat-space thermal state, a pseudoscalar has zero expectation value if parity is preserved. Therefore, the leading term in \eqref{eq:high-T-expansion} vanishes, and we have to consider the next order in the thermal expansion.

The thermal state is defined by the timelike Killing flow $\partial_t$.
The corresponding unit vector is
$e^\mu=(1,0,0)/\sqrt{1+y^2}$. In a stationary spacetime, this vector
need not be orthogonal to a family of spacelike hypersurfaces. The failure of this orthogonality is measured by its vorticity.
On the $(2+1)d$ pp-wave background without external fields, vorticity $\omega$ becomes the only pseudoscalar at the first derivative order,
\begin{equation}
    \omega=-\frac12\varepsilon^{\mu\nu\rho}
    e_\mu\nabla_\nu e_\rho
    =\frac{y}{1+y^2}.
\end{equation}

For a pseudoscalar primary ${\mathcal O}$ of dimension $\Delta_{{\mathcal O}}$,
dimensional analysis fixes this contribution to be
$b^{(\omega)}_{{\mathcal O}}\omega\widetilde\beta^{\,1-\Delta_{{\mathcal O}}}$. Substituting the local inverse temperature $\widetilde\beta=\beta\sqrt{1+y^2}$, we obtain a pseudoscalar one-point function
\begin{equation}
\langle{\mathcal O}(y)\rangle_{\mathrm{pp},\beta}
\overset{\beta\to 0}{\simeq}
\frac{b^{(\omega)}_{\mathcal O}}{\beta^{\Delta_{\mathcal O}-1}}
\frac{y}{(1+y^2)^{(1+\Delta_{\mathcal O})/2}}.
\end{equation}
As a result, the leading allowed response is odd in $y$ and contains one fewer power of temperature than the scalar response of the same dimension.

It would be interesting to extend this analysis to the critical Gross-Neveu model at large $N$ on the pp-wave background. In the Hubbard-Stratonovich formulation, the field $\sigma$ couples to the fermion bilinears $\bar\psi_i\psi_i$ and therefore must transform as a pseudoscalar. 
The non-vanishing free-fermion one-point function computed above suggests that the large-$N$ gap equation admits a nonzero, spatially varying saddle. Determining this saddle and its temperature dependence would provide a non-trivial check of the vorticity-induced response discussed above.

\section{Heavy-light-heavy correlator on a cylinder}\label{ref:HHLcylinder}
In this section, we show that the low-temperature prediction matches the analytic-bootstrap description in terms of weakly coupled particles on the cylinder. We first perform this check in free theory as a warm-up: the generalization to a generic CFT is then straightforward. Finally we comment on the Weyl rescaling and Laplace transform to obtain explicit predictions for averaged HLH three-point functions in CFT.

\subsection{Free scalar theory in (2+1)d}

In this section, we illustrate the general microscopic derivation in the simplest example of a free scalar field on a unit-radius cylinder $\mathbb{R}\times S^{2}$. The Hamiltonian of the theory is the dilatation operator, and the spatial $SO(3)$ rotation generators obey
\begin{equation}
    [{J}_z,{J}_\pm]=\pm {J}_\pm,\qquad [{J}_+,{J}_-]=2 {J}_z.
\end{equation} 

Single-particle states furnish spin-$j$ representations, and the large-spin sector is naturally organized around highest-weight states. Since the derivative $\partial_+=(\partial_1+i\partial_2)/\sqrt{2}$ increases both the scaling dimension and the spin by one unit, it leaves the twist invariant. Consequently, the lowest-twist heavy operators are obtained by taking products of highest-weight constituents,
\begin{equation}
{:}\partial_+^{j_1}\phi\,\partial_+^{j_2}\phi\cdots\partial_+^{j_N}\phi{:} ,
\qquad
\tau=\frac{N}{2}.
\end{equation}
The remaining states in each spin-$j$ multiplet are generated by repeated action of the lowering operator $J_-$, with the magnetic quantum number $m=-j,...,j$.  Namely,
\begin{equation}
\phi_{j,m}
\propto
J_-^{j-m}
\partial_+^{\,j}\phi.
\end{equation}
For a generic scalar primary $\mathcal P$, which is discussed in the next section, one would also have descendants obtained by acting with powers of the Laplacian, $\Box^s\mathcal P$. These play the same role as the longitudinal oscillator excitations discussed in the previous section.

In order to make contact with the pp-wave spectrum, we introduce a new set of quantum numbers,
\begin{equation}
\ell=|m|,\qquad n=j-|m|,
\end{equation}
and split these descendants into two groups:
\begin{equation}
\phi^{(+)}_{\ell,n}
\propto
J_-^{n}
\partial_+^{\ell+n}\phi,\qquad \phi^{(-)}_{\ell,n}
\propto
J_+^{n}
\partial_-^{\ell+n}\phi.
\end{equation}
The corresponding scaling dimensions and twists are as follows:
\begin{equation}
\Delta_{\phi^{(+)}_{\ell,n}}
=
\ell+n+\frac{1}{2},
\qquad
\tau_{\phi^{(+)}_{\ell,n}}
=
n+\frac12,
\end{equation}
and
\begin{equation}
\Delta_{\phi^{(-)}_{\ell,n}}
=
\ell+n+\frac{1}{2},
\qquad
\tau_{\phi^{(-)}_{\ell,n}}
=
2\ell+n+\frac12,
\end{equation}

Since the theory is free, heavy operators admit a simple Fock-space description. A generic heavy operator is specified by the occupation numbers $N^{(\pm)}_{\ell,n}$ of the single-particle modes,
\begin{equation}
    {:}\prod_{\ell,n\geq0}
\big(\phi^{(+)}_{\ell,n}\big)^{\,N^{(+)}_{\ell,n}}\big(\phi^{(-)}_{\ell,n}\big)^{\,N^{(-)}_{\ell,n}}{:},
\end{equation}
where we assume a bookkeeping convention $N^{(-)}_{0,n}\equiv0$.
The corresponding total spin projection and twist are
\begin{equation}
J_z
=
\sum_{\ell,n}
\ell \big(N^{(+)}_{\ell,n}-N^{(-)}_{\ell,n}\big),
\qquad
\tau
=
\sum_{\ell,n}\left[
\left(
n+\frac12
\right)
N^{(+)}_{\ell,n}+\left(2\ell+
n+\frac12
\right)
N^{(-)}_{\ell,n}\right].
\end{equation}
At $t=0$, the scalar field admits the spherical harmonic expansion
\begin{equation}
\phi(0,\hat\Omega)
=
\sum_{j=0}^{\infty}
\sum_{m=-j}^{j}
\frac{1}{\sqrt{2j+1}}
\left(
b_{j,m}Y_{j,m}(\hat\Omega)
+
b_{j,m}^\dagger
Y_{j,m}^*(\hat\Omega)
\right).
\end{equation}
Introducing the operators $b^{(+)}_{\ell,n}=b_{\ell+n,\ell}$ and $b^{(-)}_{\ell,n}=b_{\ell+n,-\ell}$, the normalized heavy state is
\begin{equation}
|\tau,J_z,\{N\}\rangle
=
\prod_{\ell,n}
\frac{
\big({b^{(+)}_{\ell,n}}^\dagger\big)^{N^{(+)}_{\ell,n}}\big({b^{(-)}_{\ell,n}}^\dagger\big)^{N^{(-)}_{\ell,n}}
}{
\sqrt{N^{(+)}_{\ell,n}!N^{(-)}_{\ell,n}!}
}
|0\rangle.
\end{equation}

As a probe, we consider the simplest scalar operator with a non-vanishing diagonal matrix element, $\langle\tau,J_z,\{N\}|
{:}\phi^2{:}
|\tau,J_z,\{N\}\rangle$.
For fixed values of $(\tau,J_z)$, the spectrum is highly degenerate. Rather than considering a particular heavy operator, it is therefore natural to study the microcanonical average over all states with the same conserved charges,
\begin{equation}
\langle\tau,J_z|
{:}\phi^2{:}
(\hat\Omega)
|\tau,J_z\rangle
=
\sum_{\ell,n}
\Big(\langle N^{(+)}_{\ell,n}\rangle_{\tau,J_z}+\langle N^{(-)}_{\ell,n}\rangle_{\tau,J_z}\Big)
\frac{
|Y_{\ell+n,\,\ell}(\hat\Omega)|^2
}{
\ell+n+\frac12
}.
\end{equation}

The microcanonical occupation numbers are most conveniently obtained by first introducing the grand-canonical partition function
\begin{equation}
\mathcal Z(q,p)
=
\operatorname{Tr}
q^\tau
p^{J_z}
=
\prod_{n\geq0}\bigg[\prod_{\ell\geq0}
\frac{
1
}{
1-w^{(+)}_{\ell,n}
}\prod_{\ell'\geq1}
\frac{
1
}{
1-w^{(-)}_{\ell',n}
}\bigg],
\end{equation}
where we introduced fugacities
\begin{equation}
q=e^{-\beta},
\qquad
p=e^{-\mu},
\end{equation}
with $\beta$ and $\mu$ conjugate to the twist and spin, respectively, and weights 
\begin{equation}
w^{(+)}_{\ell,n}=q^{n+\frac12}p^{\ell},\qquad w^{(-)}_{\ell,n}=q^{2\ell+n+\frac12}p^{-\ell}.
\end{equation} 
Denoting $\mathcal{C}_{q^\tau p^{J_z}}[f]$ for extracting the
corresponding microcanonical coefficient, or equivalently an inverse Laplace transform, the
average occupation is
\begin{equation}
\langle\tau,J_z| N^{(\pm)}_{\ell,n}|\tau,J_z\rangle
=
\frac{
\mathcal{C}_{q^\tau p^{J_z}}\left[
w^{(\pm)}_{\ell,n}
\big(1-w^{(\pm)}_{\ell,n}\big)^{-1}
\,
\mathcal Z(q,p)
\right]}{\mathcal{C}_{q^\tau p^{J_z}}\left[
\mathcal Z(q,p)
\right]}.
\end{equation}

In the thermodynamic limit, where both $\tau$ and $J_z$ become large, the coefficient extraction is governed by a saddle point. The occupation numbers reduce to the familiar Bose-Einstein distribution,
\begin{equation}
\langle\tau,J_z| N^{(+)}_{\ell,n}|\tau,J_z\rangle
\simeq
\frac{
1
}{
e^{\beta_*(n+\frac12)+\mu_*\ell}-1
},\qquad \langle \tau,J_z |N^{(-)}_{\ell,n}|\tau,J_z\rangle
\simeq
\frac{
1
}{
e^{\beta_*(n+\frac12)+(2\beta_*-\mu_*)\ell}-1
}
\end{equation}
where the saddle-point values $\beta_*(\tau,J)$ and $\mu_*(\tau,J)$ can be determined exactly.

Depending on the relative scaling of the two conserved charges, the saddle admits two qualitatively different regimes. When the twist is parametrically smaller than $\sqrt J_z$, the system is effectively at low temperature,
\begin{equation}
1\ll\tau\ll\sqrt J_z,
\qquad
\beta_*
\simeq
4\log\!\left(
\frac{\sqrt J_z}{2\tau}
\right),
\qquad
\mu_*
\simeq
\frac{2\tau}{J_z}.
\end{equation}
Conversely, when the twist dominates over $\sqrt J_z$, the effective temperature becomes large,
\begin{equation}
\sqrt{J_z}\ll\tau\ll J_z,
\qquad
\beta_*
\simeq
\left(
\frac{\zeta(3)J_z}{\tau^2}
\right)^{1/3},
\qquad
\mu_*
\simeq
\left(
\frac{\zeta(3)\tau}{J_z^2}
\right)^{1/3}.
\end{equation}
The crossover between the two behaviors occurs for
\begin{equation}
\tau\sim\sqrt J_z,
\end{equation}
corresponding to an effective temperature of order one. At any effective temperature in the fast-rotation limit, $\beta_*\gg\mu_*$. For modes in the large-spin sector $\ell\sim1/\mu_*$, negative-spin occupations are exponentially suppressed. We therefore retain the positive-spin contribution when extracting the leading asymptotics of this sector.

To compare with the pp-wave description, we isolate the large-spin sector by splitting the sum over $\ell$ into a finite part and an asymptotic part,
\begin{equation}
\langle\tau,J_z|{:}\phi^2{:}(\theta,\varphi)|\tau,J_z\rangle
=
\sum_n
\left[
\sum_{\ell=0}^{L}
+
\sum_{\ell=L+1}^{J\rightarrow\infty}
\right]
\frac{
|Y_{\ell+n,\ell}(\theta,\varphi)|^2
}{
(\ell+n+\frac12)
\left(
e^{\beta_*(n+\frac12)+\mu_*\ell}-1
\right)
},
\end{equation}
where $L$ is kept fixed while the second sum is taken to the continuum limit.
Introducing the scaling variables
\begin{equation}
\epsilon_*=\frac{\mu_*}{\beta_*},
\qquad
k=2\epsilon_*\ell,
\qquad
\theta
=
\frac{\pi}{2}
-
\sqrt{2\epsilon_*}\,y,
\end{equation}
the large-$\ell$ asymptotics of the spherical harmonics becomes
\begin{equation}
Y_{\ell+n,\ell}(\theta,\varphi)
\longrightarrow
\frac{e^{i\ell\varphi}}
{\sqrt{2\pi}}
(2\epsilon_*)^{-1/4}
\varphi_{k,n}(y),
\end{equation}
where $\varphi_{k,n}(y)$ are precisely the harmonic oscillator wave functions that appear in the scalar quantization on the pp-wave.
Replacing the large-$\ell$ sum by an integral, we find
\begin{multline}
\langle\tau,J_z|{:}\phi^2{:}(y)|\tau,J_z\rangle
=
\sum_n
\Bigg[
\sum_{\ell=0}^{L}
\frac{
|Y_{\ell+n,\ell}|^2
}{
(\ell+n+\frac12)
\left(
e^{\beta_*(n+1/2)+\mu_*\ell}-1
\right)
}
\\
+
(2\epsilon_*)^{-1/2}
\int_{2\epsilon_*(L+1)}^\infty
\frac{dk}{2\pi k}
\frac{
|\varphi_{k,n}(y)|^2
}{
e^{\beta_*(k/2+n+1/2)}-1
}
\Bigg]+O(1).
\end{multline}

Since $\epsilon_*\to0$ in the large-spin limit while $L$ remains fixed, the first term in the brackets is subleading, and we obtain
\begin{equation}
\langle\tau,J_z|{:}\phi^2{:}(y)|\tau,J_z\rangle
=
(2\epsilon_*)^{-1/2}\sum_n
\int_0^\infty
\frac{dk}{2\pi k}
\frac{
|\varphi_{k,n}(y)|^2
}{
e^{\beta_*(k/2+n+1/2)}-1
}
+
O(1).
\end{equation}
The right-hand side is precisely the thermal one-point function on the $(2+1)$-dimensional pp-wave background evaluated at the effective temperature $\beta_*^{-1}$. We therefore recover the correspondence
\begin{equation}
\langle\tau,J_z|{:}\phi^2{:}(y)|\tau,J_z\rangle
\simeq
(2\epsilon_*)^{-1/2}\langle:\!\phi^2\!:(y)\rangle_{\rm pp,\beta_*},
\end{equation}
which establishes the equivalence between heavy-light-heavy correlators on the cylinder and thermal one-point functions on the emergent pp-wave geometry.

\subsection{The lightcone limit case}
We now extend the previous discussion to a generic conformal field theory. The analytic bootstrap predicts that, in the large-spin and low-twist regime,  the leading thermodynamics admits an effectively weakly coupled description in terms of descendants of a lightest scalar primary $\mathcal{A}$ with twist $\tau_{\rm min}$. Its descendants are generated by acting with spin-raising, spin-lowering and trace operators,
\begin{equation}
\mathcal A_{\ell,n,s}
\propto
\Box^s
J_-^n
\partial_+^{\ell+n}
\mathcal A.
\end{equation}
Within the dilute gas approximation, generic heavy operators are obtained by populating these single-particle modes,
\begin{equation}
{:}\prod_{\ell,n,s\geq0}
\mathcal A_{\ell,n,s}^{\,N_{\ell,n,s}}{:} ,
\end{equation}
with total charges
\begin{equation}
J_z=\sum_{\ell,n,s}\ell N_{\ell,n,s},
\qquad
\tau=\sum_{\ell,n,s}
(\tau_{\rm min}+n+2s)
N_{\ell,n,s}.
\end{equation}
As in the free scalar example, descendants with negative magnetic quantum number are subleading in the large positive-spin limit and can be neglected from the beginning.

The averaged heavy-light-heavy correlator is defined by averaging over all states with fixed $(\tau,J_z)$. Introducing the grand-canonical partition function and extracting the corresponding microcanonical coefficients, one finds that the occupation numbers are governed by the same saddle-point analysis discussed previously. In the low-temperature regime,
\begin{equation}
1\ll\tau\ll\sqrt J_z,
\end{equation}
the Bose-Einstein distribution reduces to the Boltzmann form
\begin{equation}
\langle \tau,J_z |N_{\ell,n,s}|\tau,J_z\rangle
\simeq
e^{-\beta_*(\tau_{\min}+n+2s)-\mu_*\ell}.
\end{equation}
The leading thermodynamic contribution comes from $n=s=0$ modes, giving
\begin{equation}
\log \mathcal{Z}\simeq \frac{e^{-\beta\tau_{\min}}}{\mu},
\end{equation}
and the saddle-point equations give
\begin{equation}
\beta_*
\simeq
\frac{2}{\tau_{\min}}
\log\!\left(
\frac{\tau_{\min}\sqrt J_z}{\tau}
\right),
\qquad
\mu_*
\simeq
\frac{\tau}{\tau_{\min}J_z}.
\end{equation}

However, the minimal-twist primary $\mathcal A$ that governs thermodynamics need not couple diagonally to the probe operator $\mathcal{O}$. We therefore distinguish $\mathcal A$
from the lightest primary $\mathcal P$ that satisfies\footnote{In particular, for $\mathcal O={:}\phi^4{:}$ in the free
scalar theory, Wick's theorem gives $C_{\phi\phi\mathcal O}=0$,
while the lightest diagonally coupled scalar primary is
$\mathcal P={:}\phi^2{:}$.
As a result, the twist $\tau_{\min}=1/2$ controls thermodynamics, whereas
$\tau_{\mathcal P}=1$ defines the leading Boltzmann suppression and the kernel.
We thank Sridip Pal for bringing this point to our attention.}
\begin{equation}
C_{\mathcal P\mathcal P\mathcal O}\neq0, \qquad \tau_{\mathcal P}\geq\tau_{\min}.
\end{equation}
Since descendants with additional twist are exponentially suppressed, the leading contribution comes entirely from the lightest primary trajectory,
\begin{equation}
\langle\tau,J_z|\mathcal O(\hat\Omega)|\tau,J_z\rangle
\simeq
e^{-\beta_*\tau_{\mathcal P}}
\sum_{\ell}
e^{-\mu_*\ell}
\,
\mathcal K_{\ell}(\hat\Omega)+\dots,
\end{equation}
where we introduce the kernel
\begin{equation}
    \mathcal K_\ell(\hat\Omega)
    =
    \frac{
        \langle\ell|
        \mathcal O(\hat\Omega)
        |\ell\rangle
    }{
        \langle\ell|\ell\rangle
    },
    \qquad
    |\ell\rangle
    \propto
    \partial_+^\ell\mathcal P(0)|0\rangle .
\end{equation}
The ellipsis denotes the subleading contributions of many-particle states, transverse descendants, and higher-twist primary families.

The remaining ingredient is therefore the diagonal matrix element of the probe operator between large-spin descendants. Conformal symmetry completely fixes this quantity. Introducing the generating functions for $\langle\ell|
\mathcal O
|\ell\rangle$ and $\langle\ell|
\ell\rangle$, and extracting the appropriate coefficients, we find the $\ell$-descendant kernel
\begin{equation}
\mathcal K_\ell(\theta)
=
\frac{C_{\mathcal P\mathcal P\mathcal O}}{C_{\mathcal P}}
\frac{\ell!}{(\tau_{\mathcal P})_\ell}
\sum_{m=0}^{\ell}
\frac{
(\tau_{\mathcal P}-\Delta_{\mathcal O}/2)_{\ell-m}
}{
(\ell-m)!
}
\frac{
(\Delta_{\mathcal O}/2)_m^2
}{
(m!)^2
}
\sin^{2m}\theta,
\label{eq:generic-kernel}
\end{equation}
where axial symmetry removes the dependence
on the azimuthal angle.

We focus on the regime $\Delta_{\mathcal O}>1$, for which the large-spin kernel is dominated by the terms $m=O(\ell)$ and admits the pp-wave limit.
To compare with the pp-wave computation, we introduce the scaling limit
\begin{equation}
k=2\epsilon_*\ell,
\qquad
\theta=\frac{\pi}{2}-\sqrt{2\epsilon_*}\,y,
\qquad
\epsilon_*=\frac{\mu_*}{\beta_*},
\qquad
\ell\rightarrow\infty,
\end{equation}
while keeping $k$ and $y$ fixed. Remarkably, the kernel admits a smooth continuum limit,
\begin{multline}
\mathcal K_\ell
\left(
\frac{\pi}{2}-\sqrt{2\epsilon_*}\,y
\right)
\longrightarrow
(2\epsilon_*)^{1-\Delta_{\mathcal O}/2}
\frac{C_{\mathcal P\mathcal P\mathcal O}}{C_{\mathcal P}}
\frac{
\Gamma(\tau_{\mathcal P})
\Gamma(\Delta_{\mathcal O}-1)
}{
\Gamma(\Delta_{\mathcal O}/2)^2
\Gamma(\tau_{\mathcal P}+\Delta_{\mathcal O}/2-1)
}
\\
\times
k^{\Delta_{\mathcal O}/2-1}
{}_1F_1
\!\left(
\Delta_{\mathcal O}-1;
\tau_{\mathcal P}+\frac{\Delta_{\mathcal O}}2-1;
-ky^2
\right),
\end{multline}
which is precisely the one-particle kernel obtained in the pp-wave quantization.
Replacing the large-spin sum by a continuum integral then gives
\begin{multline}
\langle\tau,J_z|\mathcal O(y)|\tau,J_z\rangle
\simeq (2\epsilon_*)^{-\Delta_{\mathcal O}/2}
e^{-\beta_*\tau_{\mathcal P}}
\frac{C_{\mathcal P\mathcal P\mathcal O}}{C_{\mathcal P}}
\frac{
\Gamma(\tau_{\mathcal P})
\Gamma(\Delta_{\mathcal O}-1)
}{
\Gamma(\Delta_{\mathcal O}/2)^2
\Gamma(\tau_{\mathcal P}+\Delta_{\mathcal O}/2-1)
}
\\
\times
\int_0^\infty
dk\,
e^{-\beta_*k/2}
k^{\Delta_{\mathcal O}/2-1}
{}_1F_1
\!\left(
\Delta_{\mathcal O}-1;
\tau_{\mathcal P}+\frac{\Delta_{\mathcal O}}2-1;
-ky^2
\right),
\end{multline}
whose evaluation immediately reproduces the universal low-temperature one-point function derived above. We therefore obtain
\begin{equation}
\langle\tau,J_z|\mathcal O(y)|\tau,J_z\rangle
\simeq
(2\epsilon_*)^{-\Delta_{\mathcal O}/2}\langle\mathcal O(y)\rangle_{\rm pp,\beta_*},\label{eq:cyl-pp-wave}
\end{equation}
establishing the correspondence between heavy-heavy-light correlators on the cylinder and thermal one-point functions on the emergent pp-wave geometry for a generic conformal field theory. Substituting the saddle point values $\epsilon_*$ and $\beta_*$, and $y=(\pi/2-\theta)/\sqrt{2\epsilon_*}$, gives the universal asymptotics in the small-twist regime $\tau\ll \sqrt{J_z}$,
\begin{multline}
    \langle\tau,J_z|\mathcal O(\theta)|\tau,J_z\rangle\simeq
    \left(
        \frac{\tau_{\min}}{\tau}
    \right)^{\Delta_{\mathcal O}/2-2\tau_{\mathcal P}/\tau_{\min}}
    J_z^{\Delta_{\mathcal O}/2-\tau_{\mathcal P}/\tau_{\min}}\frac{C_{\mathcal P\mathcal P\mathcal O}}{C_{\mathcal P}}
\\
    \times 
    \frac{
\Gamma(\tau_{\mathcal P})
\Gamma(\Delta_{\mathcal O}-1)
}{
\Gamma(\Delta_{\mathcal O}/2)
\Gamma(\tau_{\mathcal P}+\Delta_{\mathcal O}/2-1)
}
    {}_2F_1\!\left(
        \Delta_{\mathcal O}-1,
        \frac{\Delta_{\mathcal O}}{2};
        \tau_{\mathcal P}+\frac{\Delta_{\mathcal O}}{2}-1;
        -\frac{\tau_{\min}J_z}{\tau}\left(\theta-\frac{\pi}{2}\right)^2
    \right).
\end{multline}
Finally, we note that the relation is applicable at any effective temperature $\beta_*$.  Substituting the high-temperature saddle produces the large-twist asymptotic behavior of HLH correlators,
\begin{equation}
   \langle\tau,J_z|\mathcal O(\theta)|\tau,J_z\rangle\simeq \frac{b_\mathcal{O}(\tau J_z)^{\Delta_{\mathcal O}/6}}{\left[1+\frac{J_z}{2\tau}\left(\theta-\frac{\pi}{2}\right)^2\right]^{\Delta_{\mathcal O}/2}},\qquad\sqrt{J_z}\ll\tau\ll J_z.
\end{equation}

Some comments are in order. Extending beyond the regime $\Delta_{\mathcal O}>1$ exhibits different equatorial asymptotics. In the ultralight regime with ${1}/{2}<\Delta_{\mathcal O}<1$, the leading kernel is controlled by $m\ll\ell$, retaining a non-zero $\theta$-profile away from the equatorial region. At $\Delta_{\mathcal O}=1$ and $\tau_{\mathcal{P}}>1/2$, a broad range of terms with $1\ll m\ll \ell$ contribute to the kernel, and the correlator develops a logarithmic enhancement. However, when $\tau_{\mathcal{P}}=1/2$, the primary $\mathcal{P}$ saturates the scalar unitarity bound, and the kernel collapses to $    \mathcal{K}_\ell(\theta)\sim|Y_{\ell,\ell}(\theta)|^2/(\ell+1/2)$.

\subsection{Weyl rescaling and Laplace transform}
\label{sec:weyl-laplace}

To obtain an averaged
heavy-light-heavy matrix element at fixed twist and angular momenta, we have to express the corresponding one-point function on the cylinder using the Weyl map, and then extract the contribution with the corresponding charges.
Let us illustrate these two steps on the example of a $(3+1)$-dimensional theory with two large
angular momenta.

We consider a unit-radius cylinder and choose angular coordinates such that
\begin{equation}
 ds_{\rm cyl}^{2}=-dt^{2}+d\vartheta^{2}
 +\sin^{2}\vartheta\,d\varphi_{1}^{2}
 +\cos^{2}\vartheta\,d\varphi_{2}^{2},
 \qquad 0\leq\vartheta\leq\frac{\pi}{2},
\end{equation}
where $J_i$ generates translations of $\varphi_i$. The corresponding ensemble is
\begin{equation}
 \mathcal{Z}(q,p_1,p_2)
 =\operatorname{Tr}
 q^\tau p_1^{J_1}p_2^{J_2},\qquad p_i=e^{-\mu_i}.
\end{equation}
The variables conjugate to $(\tau,J_1,J_2)$ are
$(\beta,\mu_1,\mu_2)$, and the corresponding  angular velocities are $\Omega_i=1-\mu_i/\beta$. In the rotating coordinates, the pp-wave limit gives
\begin{equation}
ds_{\rm cyl}^2\simeq\gamma(\vartheta)ds_{{\rm pp}_2}^2.
\end{equation}
The Weyl factor can be derived directly from the norm of the Killing vector, 
\begin{equation}
K^\mu\partial_\mu=\partial_t+\Omega_1\partial_{\varphi_1}
+\Omega_2\partial_{\varphi_2}.
\end{equation}
Denoting $\epsilon_i=\mu_i/\beta$, we find
\begin{equation}
 -g_{\rm cyl}(K,K)=1-\Omega_1^2\sin^2\vartheta-\Omega_2^2\cos^2\vartheta
   =\gamma(\vartheta)+O(\epsilon_i^2),
\end{equation}
where
\begin{equation}
 \gamma(\vartheta)
 ={2}
\left(\epsilon_1\sin^2\vartheta+\epsilon_2\cos^2\vartheta\right).
 \label{eq:pp2-hlh-weyl}
\end{equation}
As a result, for a scalar primary probe
$\mathcal O$ of dimension $\Delta_{\mathcal O}$, we obtain
\begin{equation}
 \langle\mathcal O(\vartheta)\rangle_{\beta,\mu_1,\mu_2}
 \simeq\gamma(\vartheta)^{-\Delta_{\mathcal O}/2}
 \langle\mathcal O\rangle_{{\rm pp}_2,\beta}.
 \label{eq:correlators}
\end{equation}
Although the pp-wave isometries imply that $\langle\mathcal O\rangle_{{\rm pp}_2,\beta}$ is
spatially constant, the cylinder one-point function retains the angular
dependence through the Weyl factor.

Let $\rho(\tau,J_1,J_2)$ be the density of states on the cylinder, and define the averaged diagonal HLH matrix element,
\begin{equation}
\langle\tau,J_1,J_2|{\mathcal O}(\vartheta)|\tau,J_1,J_2\rangle
 =\frac{1}{\rho(\tau,J_1,J_2)}
 \sum_{i}
 \langle \tau,J_1,J_2,i|\mathcal O(\vartheta)|\tau,J_1,J_2,i\rangle.
\end{equation}
Using the inverse Laplace transform, or equivalently, the coefficient extraction, and substituting \eqref{eq:correlators}, this becomes
\begin{equation}
\langle\tau,J_1,J_2|{\mathcal O}(\vartheta)|\tau,J_1,J_2\rangle
 =\frac{\mathcal{C}_{q^\tau p_1^{J_1}p_2^{J_2}}\left[\gamma(\vartheta)^{-\Delta_{\mathcal O}/2}
 \langle\mathcal O\rangle_{{\rm pp}_2,\beta}\mathcal{Z}(q,p_1,p_2)\right]}{\mathcal{C}_{q^\tau p_1^{J_1}p_2^{J_2}}\left[\mathcal{Z}(q,p_1,p_2)\right]}.
\end{equation}

In the limit of large charges, the Laplace transform admits a saddle point approximation. Using the expression for $\gamma(\vartheta)$, we obtain
\begin{equation}
 \langle\tau,J_1,J_2|\mathcal O(\vartheta)|\tau,J_1,J_2\rangle
 \simeq\left({2}
{\epsilon_1}_*\sin^2\vartheta+2{\epsilon_2}_*\cos^2\vartheta\right)^{-\Delta_{\mathcal O}/2}
 \langle\mathcal O\rangle_{{\rm pp}_2,\beta_*}.
\end{equation}
The saddle can be found analytically in the low- and high-twist regimes. Indeed, the $\beta_*\gg1$ saddle is
\begin{equation}
\tau\ll(J_1J_2)^{1/3},\qquad\beta_*
\simeq
\frac{3}{\tau_{\min}}
\log\!\left(
\frac{\tau_{\min}(J_1J_2)^{1/3}}{\tau}
\right),
\qquad
{\mu_i}_*
\simeq
\frac{\tau}{\tau_{\min}J_i},
\end{equation}
and 
the $\beta_*\ll1$ saddle is
\begin{equation}
(J_1J_2)^{1/3}\ll\tau\ll\min\{J_1,J_2\},\qquad\beta_*
\sim
\left(
\frac{(J_1J_2)^{1/3}}{\tau}
\right)^{3/4},
\qquad
{\mu_i}_*
\simeq
\frac{\beta_*\tau}{J_i}.
\end{equation}
As a result, the average HLH correlator on a 3+1 dimensional cylinder with two large angular momenta scales as
\begin{equation}
\left\langle\tau,J_1,J_2| \mathcal{O}(\vartheta) |\tau,J_1,J_2\right\rangle
\sim
\begin{cases}
\displaystyle
\frac{\left(
\frac{\tau_{\min}}
{\tau}
\right)^{\Delta_{\mathcal{O}}/2-3\tau_{\mathcal{P}}/\tau_{\min}}(J_1J_2)^{\Delta_{\mathcal{O}}/2-\tau_{\mathcal{P}}/\tau_{\min}}
}{\left(J_1\cos^2\vartheta+J_2\sin^2\vartheta
\right)^{\Delta_{\mathcal{O}}/2}},
&
\displaystyle
\tau \ll (J_1J_2)^{1/3},
\\[1.6em]
\displaystyle
\frac{(\tau J_1J_2)^{\Delta_{\mathcal{O}}/4}
}{(J_1\cos^2\vartheta+J_2\sin^2\vartheta
)^{\Delta_{\mathcal{O}}/2}},
&
\displaystyle
\tau \gg (J_1J_2)^{1/3}.
\end{cases}
\end{equation}

\acknowledgments

We are grateful to Zohar Komargodski, Jeremy Mann, Sridip Pal, Leonardo Rastelli, and Francesco Russo. We especially thank Zohar Komargodski for many crucial discussions at different stages of the project. We thank Nathan Benjamin, Zohar Komargodski, Francesco Mangialardi, Jeremy Mann, and Sridip Pal for useful comments on the manuscript.

\appendix

\section{Spectrum on pp-wave}\label{eq:Hinppwave}
In this appendix we describe in more detail the Hilbert space associated with the pp-wave background and derive the spectrum of the corresponding Hamiltonian. The main result is that every conformal family of the original CFT gives rise to an infinite tower of pp-wave states whose energies are completely determined by the twist of the primary together with a set of harmonic oscillator excitations.

\subsection{A single large spin}
\label{app:single-large-spin}

We consider the single-large-spin pp-wave geometry
\[
ds_{\rm pp}^2
=
-\left(1+\sum_{\alpha=1}^{d-2}y_\alpha^2\right)dt^2
+2\,dt\,dx
+\sum_{\alpha=1}^{d-2}dy_\alpha^2.
\]
The pp-wave vacuum has vanishing energy,
\begin{equation}
    H_{\rm pp}\lvert {\Omega}\rangle=0 .
    \label{eq:app-single-vacuum}
\end{equation}

To make the conformal equivalence to the flat space explicit, it is
convenient to introduce the shifted null coordinate
${x'}=x-t/2$. In terms of $(t,x',\vec y)$ the pp-wave metric
takes the form
\begin{equation}
    ds_0^2
    =
    -\sum_{\alpha=1}^{d-2}y_\alpha^2dt^2
    +2\,dt\,dx'
    +\sum_{\alpha=1}^{d-2}dy_\alpha^2 .
    \label{eq:app-single-aux-metric}
\end{equation}
Using the following coordinate transformation,
\begin{equation}
    \xi=\tan t,
    \qquad
    \eta=x'-\frac{y^2}{2}\tan t,
    \qquad
    w_\alpha=\frac{y_\alpha}{\cos t},
    \label{eq:app-single-flat-map}
\end{equation}
the metric becomes
\begin{equation}
    ds_0^2
    =
    \cos^2 t
    \left(
        2\,d\xi\,d\eta
        +\sum_{\alpha=1}^{d-2}dw_\alpha^2
    \right),
    \label{eq:app-single-flat-metric}
\end{equation}
manifesting the conformal equivalence of the pp-wave to a flat space.

Let $H_0$ denote the generator of translations in $t$ at fixed
${x'}$. Under the conformal map above it is mapped to
\begin{equation}
    H_0=P_\xi-K_\eta ,
    \label{eq:app-single-H0}
\end{equation}
which is conjugate to the light-cone twist operator,
\begin{equation}
    \mathcal{U} H_0 \mathcal{U}^{-1}=\mathcal{T},
    \qquad
    \mathcal{T}=D+2M_{\xi\eta},
    \qquad
    \mathcal{U}=
    \exp\left[
        \frac{i\pi}{4}\left(P_\xi+K_\eta\right)
    \right].
    \label{eq:app-single-twist}
\end{equation}
As a result, the spectrum of $H_0$ is determined by the
light-cone twist spectrum of the original conformal field theory.

Let $\mathcal{P}$ denote a scalar primary with twist $\tau_{\mathcal{P}}$. It
defines a lowest-energy state through the operator-state
correspondence,
\begin{equation}
    \mathcal{U}^{-1}\lvert \mathcal{P}\rangle
    =
    \lim_{t_E\to-\infty}
    e^{-t_E\tau_{\mathcal{P}}}
    \mathcal{P}(-it_E,0,\vec 0)\lvert {\Omega}\rangle
    =
    2^{\tau_{\mathcal{P}}}
    \mathcal{P}_{\rm flat}(i,0,\vec 0)\lvert {\Omega}\rangle ,
    \label{eq:app-single-primary-ket}
\end{equation}
while the corresponding bra state is
\begin{equation}
    \langle \mathcal{P}\rvert \mathcal{U}
    =
    2^{\tau_{\mathcal{P}}}
    \langle {\Omega}\rvert
    \mathcal{P}_{\rm flat}(-i,0,\vec 0).
    \label{eq:app-single-primary-bra}
\end{equation}
For the scalar primary under consideration, we define
\begin{equation}
    \mathcal{P}_{\rm flat}(\xi,\eta,\vec w)
    \equiv
    (\cos t)^{\tau_{\mathcal P}}
    \mathcal{P}(t,{x'},\vec y).
    \label{eq:app-single-Pflat}
\end{equation}
Since $\mathcal{T}\lvert \mathcal{P}\rangle=\tau_\mathcal{P}\lvert \mathcal{P}\rangle$, it follows
immediately that
\begin{equation}
    H_0 \mathcal{U}^{-1}\lvert \mathcal{P}\rangle
    =
    \tau_\mathcal{P} \mathcal{U}^{-1}\lvert \mathcal{P}\rangle .
    \label{eq:app-single-primary-energy}
\end{equation}
Thus, every scalar primary gives rise to the lowest energy
state of a pp-wave multiplet, with energy equal to its twist.

The remaining states are obtained algebraically by acting with
translation generators. Their action follows from 
\begin{equation}
[D,P_\mu]=P_\mu,
\qquad
[\mathcal{T},P_\eta]=0,
\qquad
[\mathcal{T},P_\alpha]=P_\alpha,
\qquad
[\mathcal{T},P_\xi]=2P_\xi.
\label{eq:app-single-descendants}
\end{equation}
Thus, $P_\eta$ moves along an orbit of fixed twist, while $P_\alpha$
and $P_\xi$ raise the twist by one and two units, respectively.

To make the transverse oscillator structure manifest, we define
\begin{equation}
    Q_\alpha=P_\alpha,
    \qquad
    R_\alpha=i[P_\alpha,K_\eta].
    \label{eq:app-single-QR}
\end{equation}
Their algebra is
\begin{equation}
    [Q_\alpha,R_\beta]
    =
    i\delta_{\alpha\beta}P_\eta .
    \label{eq:app-single-Heisenberg}
\end{equation}
Restricting to a sector of fixed positive light-cone momentum,
\begin{equation}
    P_\eta=k,
    \qquad
    k>0,
    \label{eq:app-single-k}
\end{equation}
this reduces to the standard Heisenberg algebra. Introducing
\begin{equation}
    b_{k,\alpha}
    =
    \frac{Q_\alpha+iR_\alpha}{\sqrt{2k}},
    \qquad
    b_{k,\alpha}^{\dagger}
    =
    \frac{Q_\alpha-iR_\alpha}{\sqrt{2k}},
    \label{eq:app-single-oscillators}
\end{equation}
we obtain
\begin{equation}
    [b_{k,\alpha},b_{k,\beta}^{\dagger}]
    =
    \delta_{\alpha\beta},
    \qquad
    [H_0,b_{k,\alpha}^{\dagger}]
    =
    b_{k,\alpha}^{\dagger}.
    \label{eq:app-single-oscillator-algebra}
\end{equation}
Therefore, the transverse descendants organize into harmonic
oscillator excitations. Similarly, the action of $P_\xi$ generates
longitudinal excitations separated by two units of $H_0$ energy.

Now we return to the physical pp-wave coordinates with $x=x'+t/2$.
At fixed $x$, we get
\[
\left.\partial_t\right|_x
=
\left.\partial_t\right|_{{x}'}
-\frac{1}{2}\left.\partial_{{x'}}\right|_{t} .
\]
Using the conventions $H_0=i\left.\partial_t\right|_{{x}'}$ and
$P_\eta=-i\left.\partial_\eta\right|_{\xi}=-i\left.\partial_{{x'}}\right|_{t}$, the  pp-wave Hamiltonian becomes
\begin{equation}
H_{\rm pp}
=
H_0+\frac{1}{2}P_\eta.
\label{eq:app-single-Hpp}
\end{equation}
Since
\begin{equation}
    [H_0,P_\eta]=0,
    \label{eq:app-single-H0Pv}
\end{equation}
the two operators can be diagonalized simultaneously.
A convenient basis is obtained by Fourier transforming along the
null direction,
\begin{equation}
    \lvert k,\vec n,s;\mathcal{P}\rangle
    =
    \frac{\mathcal{U}^{-1}}
    {\sqrt{\mathcal{N}_\mathcal{P}(k,\vec n,s)}}
    \int_{-\infty}^\infty\frac{d\lambda}{2\pi}\,
     e^{ik\lambda}
    e^{-i\lambda P_\eta}
    P_\xi^s
    \prod_{\alpha=1}^{d-2}
    P_\alpha^{\,n_\alpha}
    \lvert \mathcal{P}\rangle ,
    \qquad
    k>0 .
    \label{eq:app-single-basis}
\end{equation}
These states satisfy
\begin{equation}
    P_\eta\lvert k,\vec n,s;\mathcal{P}\rangle
    =
    k\lvert k,\vec n,s;\mathcal{P}\rangle ,
    \label{eq:app-single-Pv-eigenstate}
\end{equation}
and
\begin{equation}
    H_{\rm pp}\lvert k,\vec n,s;\mathcal{P}\rangle
    =
    \left(
        \frac{k}{2}
        +\tau_{\mathcal P}
        +n
        +2s
    \right)
    \lvert k,\vec n,s;\mathcal{P}\rangle ,
    \qquad
    n\equiv\sum_{\alpha=1}^{d-2}n_\alpha .
    \label{eq:app-single-spectrum}
\end{equation}
Thus, every conformal family gives rise to an infinite tower of
pp-wave states labeled by the continuous light-cone momentum $k$,
the transverse oscillator occupation numbers $\vec n$, and the
longitudinal excitation number $s$.
This decomposition immediately explains the structure of the low-temperature expansion discussed in the main text. The leading contribution comes from the lowest energy state of the lightest conformal family with $C_{\mathcal{P}\mathcal{P}\mathcal{O}}\neq0$, which we denote $\lvert k;\mathcal P\rangle\equiv\lvert k,\vec 0,0;\mathcal{P}\rangle$, whereas all remaining states are exponentially suppressed by additional Boltzmann factors.

\subsection{Two large spins}

We now turn to the $(3+1)$-dimensional pp-wave background with two large
angular momenta. Essentially, this geometry is obtained from the single-large-momentum
$(3+1)d$ pp-wave by passing to a uniformly rotating frame in the transverse
plane $(y_1,y_2)$. The spectrum follows directly from the one-spin result together with the angular-momentum shift induced by the
rotation.

In four dimensions, the single-spin pp-wave from the previous subsection
can be written as
\begin{equation}
    ds^2_{\mathrm{pp}}
    =
    -\bigl(1+y_1^2+y_2^2\bigr)dt^2
    +2\,dt\,dx+dy_1^2+dy_2^2 .
\end{equation}
The following change of coordinates
\begin{equation}
\begin{aligned}
y_1 &= y\cos t-u\sin t, \\
y_2 &= y\sin t+u\cos t, \\
x   &= -v-uy
\end{aligned}
\end{equation}
brings this metric to the required form,
\begin{equation}
    ds^2_{\mathrm{pp}_2}
    =
    -dt^2-2\,dt\,dv-4u\,dt\,dy+du^2+dy^2.
\end{equation}

Let us denote, respectively, the transverse rotation vector field and the
corresponding angular-momentum operator as
\begin{equation}
    M_{12}
    =
    y_1\partial_{y_2}-y_2\partial_{y_1},
    \qquad
    J_{12}\equiv -iM_{12}.
\end{equation}
At fixed $(u,v,y)$, the
coordinate transformation above gives
\begin{equation}
    \left.\partial_t\right|_{u,v,y}
    =
    \left.\partial_t\right|_{x,y_1,y_2}
    +M_{12}.
\end{equation}
As a result, the two-spin and one-spin Hamiltonians are related by
\begin{equation}
    H_{\mathrm{pp}_2}
    =
    H_{\mathrm{pp}}-J_{12}.
\end{equation}

Since $J_{12}$ acts only in the transverse plane, it
commutes with both $\mathcal{U}$ and $P_\eta$. Then, it follows that
\begin{equation}
    \mathcal{U} H_{\mathrm{pp}_2}\mathcal{U}^{-1}
    =
    \mathcal T_2+\frac{1}{2}P_\eta,
    \qquad
    \mathcal T_2
    \equiv
    D+2M_{\xi\eta}-J_{12}.
\end{equation}
The operator $\mathcal T_2$ is the two-spin twist operator. On a state
with cylinder quantum numbers $(\Delta,J_1,J_2)$, its eigenvalue is
\begin{equation}
    \tau=\Delta-J_1-J_2 .
\end{equation}

Introducing
the operators
\begin{equation}
    P_\pm=P_1\pm iP_2,
\end{equation}
which satisfy
\begin{equation}
    [J_{12},P_\pm]=\pm P_\pm ,
\end{equation}
we obtain
\begin{equation}
\begin{aligned}
    [\mathcal T_2,P_-]&=2P_-,
    &\qquad
    [\mathcal T_2,P_+]&=0, \\
    [\mathcal T_2,P_\xi]&=2P_\xi,
    &
    [\mathcal T_2,P_\eta]&=0 .
\end{aligned}
\end{equation}

Thus, the two transverse oscillator directions of the one-spin problem
reorganize into the familiar Landau-level structure. After conjugation
back to the pp-wave Hilbert space, we find that $\mathcal{U}^{-1}P_-\mathcal{U}$ raises the cyclotron
level, while $\mathcal{U}^{-1}P_+\mathcal{U}$ changes only the guiding center and therefore
does not change the energy. The longitudinal tower generated by $P_\xi$
retains the same level spacing.

A convenient basis of states is
\begin{equation}
|k,n,s,\kappa;\mathcal{P}\rangle
=
\frac{\mathcal{U}^{-1}}
{\sqrt{\mathcal N_\mathcal{P}(k,n,s,\kappa)}}
\int_{-\infty}^{\infty}
\frac{d\lambda}{2\pi}\,
e^{ik\lambda}
e^{-i\lambda P_\eta}
P_-^nP_\xi^sP_+^\kappa|\mathcal{P}\rangle,
\end{equation}
where $n,s,\kappa=0,1,2,..$.
Using the commutation relations above, we immediately find
\begin{equation}
    H_{\mathrm{pp}_2}|k,n,s,\kappa;\mathcal{P}\rangle
    =
    \left(
        \frac{k}{2}
        +\tau_{\mathcal P}
        +2n+2s
    \right)
    |k,n,s,\kappa;\mathcal{P}\rangle .
\end{equation}
Therefore, $n$ labels the Landau level, $\kappa$ labels its guiding-center
degeneracy, and $s$ labels the independent longitudinal tower.

\subsection{Lowest-Landau-level states}\label{sec:LLL-kernel}

In this subsection, we compute the diagonal matrix elements of a scalar primary operator
$\mathcal O$ between the lowest-Landau-level states. These matrix elements provide the remaining ingredient required to compute the low-temperature one-point functions in the two-spin pp-wave in section \ref{sec:two-large-spins}.

Let $\mathcal{P}$ be a scalar primary with twist $\tau_\mathcal{P}$, and set $n=s=0$. We denote the resulting lowest-Landau-level states by
\begin{equation}
    |k,\kappa;\mathcal P\rangle
    \equiv
    |k,0,0,\kappa;\mathcal{P}\rangle,
\end{equation}
with energy
\begin{equation}
    E_0(k)=\frac{k}{2}+\tau_{\mathcal{P}}.
\end{equation}

At fixed $k$, it is useful to introduce the canonically normalized
guiding-center oscillator. In terms of the oscillators $b_{k,1}$ and $b_{k,2}$ introduced above, we take
\begin{equation}
    a
    =
    \frac{1}{\sqrt{2}}
    \left(
        b_{k,1}-i b_{k,2}
    \right),
    \qquad
    [a,a^\dagger]=1 ,
\end{equation}
so that
\begin{equation}
|k,\kappa;\mathcal P\rangle
=
\frac{
    \left(a^\dagger\right)^\kappa
}{\sqrt{\kappa!}}
|k,0;\mathcal P\rangle .
\end{equation}
The normalized coherent states are
\begin{equation}
|\alpha\rangle
\equiv
e^{-|\alpha|^2/2}
\sum_{\kappa=0}^{\infty}
\frac{\alpha^\kappa}{\sqrt{\kappa!}}
|k,\kappa;\mathcal P\rangle .
\end{equation}

Rotational invariance implies that the coherent-state expectation value provides a
generating function for the diagonal matrix elements,
\begin{equation}
\label{eq:coherent-average}
\langle \alpha|\mathcal{O}(0)|\alpha\rangle
=
e^{-|\alpha|^2}
\sum_{\kappa=0}^{\infty}
\frac{|\alpha|^{2\kappa}}{\kappa!}
\langle k,\kappa;\mathcal P|\mathcal{O}(0)|k,\kappa;\mathcal P\rangle.
\end{equation}
Equivalently, writing 
\begin{equation}
|\alpha\rangle
=
\mathcal{D}(\alpha)|0\rangle,\qquad \mathcal{D}(\alpha)=e^{\alpha a^\dagger-\alpha^*a},
\end{equation}
the displacement operator $\mathcal{D}(\alpha)$ implements a guiding-center translation within the lowest Landau level. At fixed light-cone momentum $k$, it displaces the center by $\vec{y}(\alpha)$, with
\begin{equation}\label{eq:displaced}
k|\vec{y}(\alpha)|^2=|\alpha|^2.
\end{equation}
For a scalar operator, covariance under this isometry gives
\begin{equation}
\langle \alpha|\mathcal{O}(0)|\alpha\rangle
=
\langle k,0;\mathcal P|\mathcal{O}(\vec{y}(\alpha))|k,0;\mathcal P\rangle.
\end{equation}
The right hand side is precisely the scalar one-particle kernel obtained in \eqref{eq:one-part-kernel},
\begin{multline}
\langle k,0;\mathcal P|\mathcal{O}(\vec{y}(\alpha))|k,0;\mathcal P\rangle
=
\frac{
C_{\mathcal P\mathcal P\mathcal O}
}{
C_{\mathcal P}
}
\frac{
\Gamma(\tau_{\mathcal P})
\Gamma(\Delta_{\mathcal O}-1)
}{
\Gamma(\Delta_{\mathcal O}/2)^2
\Gamma(\tau_{\mathcal P}+\Delta_{\mathcal O}/2-1)
}
\\
\times
k^{\Delta_{\mathcal O}/2-1}
\,{}_1F_1
\!\left(
\Delta_{\mathcal O}-1;
\tau_{\mathcal P}+\frac{\Delta_{\mathcal O}}{2}-1;
-k|\vec{y}(\alpha)|^2
\right),\qquad \Delta_{\mathcal O}>1.
\end{multline}
Using \eqref{eq:displaced} and expanding the hypergeometric function, we immediately get
\begin{multline}
\langle \alpha|\mathcal{O}(0)|\alpha\rangle
= \frac{
C_{\mathcal P\mathcal P\mathcal O}
}{
C_{\mathcal P}
}
\frac{
\Gamma(\tau_{\mathcal P})
\Gamma(\Delta_{\mathcal O}-1)
}{
\Gamma(\Delta_{\mathcal O}/2)^2
\Gamma(\tau_{\mathcal P}+\Delta_{\mathcal O}/2-1)
}k^{\Delta_{\mathcal O}/2-1} \\
\times e^{-|\alpha|^2}
\sum_{\kappa=0}^{\infty}
\frac{
\left(
\tau_{\mathcal P}-{\Delta_{\mathcal O}}/{2}
\right)_\kappa
}{
\left(
\tau_{\mathcal P}+{\Delta_{\mathcal O}}/{2}-1
\right)_\kappa
}
\frac{|\alpha|^{2\kappa}}{\kappa!}.
\end{multline}
Comparing this with \eqref{eq:coherent-average} yields the desired result,
\begin{equation}
\label{eq:2-spin-kernel}
\langle k,\kappa;\mathcal P|
\mathcal O
|k,\kappa;\mathcal P\rangle=\frac{(\tau_{\mathcal P}-\Delta_{\mathcal O}/2)_\kappa}{(\tau_{\mathcal P}+\Delta_{\mathcal O}/2-1)_\kappa}\langle k,0;\mathcal P|
\mathcal O
|k,0;\mathcal P\rangle.
\end{equation}
\section{Conformal invariance of the pp-wave vacuum}\label{ref:conformalinvariancePpwave}
Consider the conformally flat metric
\begin{gather}
ds^2=-(1+y^2)\,dt^2+dt\,dx+dy^2 .
\end{gather}
We denote by \(H_0,N,P_\pm,D,K\) the conserved charges associated with the conformal Killing vectors

\begin{gather}
H_0=\partial_t,\qquad N=\partial_x, \qquad
P_\pm=e^{\pm it}\left(\partial_y\mp 2iy\,\partial_x\right),\qquad
D=2(x-t)\partial_x+y\partial_y,
\end{gather}
and, finally
\begin{gather}
K=-y^2\partial_t+
\left[(x-t)^2-y^4-y^2\right]\partial_x
+(x-t)y\partial_y .
\end{gather}
We can check that
\begin{gather}
P_+^\dagger=P_-,
\qquad
H_0^\dagger=H_0,\quad
N^\dagger=N,\quad
D^\dagger=D,\quad
K^\dagger=K.
\end{gather}
Let \(|\Omega\rangle\) be a state of lowest \(H_0\)-energy,
$H_0|\Omega\rangle=E_0|\Omega\rangle, 
E_0=\min\operatorname{Spec}H_0,
$
and assume in addition that this vacuum is translationally invariant along $x$, namely
$N|\Omega\rangle=0$.
We show that \(|\Omega\rangle\) is annihilated by the full conformal algebra.

The relevant commutators are
\begin{gather}
[H_0,P_-]=-P_-,
\qquad
[P_-,P_+]=-4N.
\end{gather}
The first relation and the lower bound on the spectrum imply that $
P_-|\Omega\rangle=0
$.
Consequently,
\begin{gather}
\|P_+|\Omega\rangle\|^2
=
\langle\Omega|[P_-,P_+]|\Omega\rangle
=
-4\langle\Omega|N|\Omega\rangle=0,
\end{gather}
and hence $P_+|\Omega\rangle=0.$ Next define the conformal partner of \(P_-\) by
$
K_-\equiv-[K,P_-].
$
The algebra gives
\begin{gather}
[H_0,K_-]=-K_-+P_-.
\end{gather}
Since \(P_-|\Omega\rangle=0\), the state \(K_-|\Omega\rangle\), if nonzero, would have energy \(E_0-1\). Therefore
, $K_-|\Omega\rangle=0
$. And one furthermore finds
$[P_+,K_-]=2(H_0+N-iD).$
Both operators on the left annihilate \(|\Omega\rangle\), and therefore
\begin{gather}
(E_0-iD)|\Omega\rangle=0.
\end{gather}
Since \(D\) is Hermitian whereas \(E_0\) is real, this equation is possible only if
$
E_0=0, D|\Omega\rangle=0
$. Finally,
$
[D,K]=-2iK.
$ Acting on the vacuum gives
$
D\,K|\Omega\rangle=-2iK|\Omega\rangle.
$

Because \(D\) is Hermitian, it cannot possess a nonzero eigenvector with the imaginary eigenvalue \(-2i\). Thus
$
K|\Omega\rangle=0.
$
At this point
\begin{gather}
H_0|\Omega\rangle=
N|\Omega\rangle=
D|\Omega\rangle=
K|\Omega\rangle=
P_\pm|\Omega\rangle=0.
\end{gather}

The four remaining conformal generators are obtained by taking commutators of these operators, for example

\begin{gather}
K_\pm\propto[K,P_\pm],
\qquad
R_\pm\propto[P_\pm,K_\pm].
\end{gather}

They therefore annihilate \(|\Omega\rangle\) as well. 
Thus, in a unitary theory, a lowest-\(H_0\) state with vanishing null momentum \(N\) is necessarily invariant under the full conformal group.

\bibliography{Draft}

@article{Hellerman:2015nra,
    author = "Hellerman, Simeon and Orlando, Domenico and Reffert, Susanne and Watanabe, Masataka",
    title = "{On the CFT Operator Spectrum at Large Global Charge}",
    eprint = "1505.01537",
    archivePrefix = "arXiv",
    primaryClass = "hep-th",
    doi = "10.1007/JHEP12(2015)071",
    journal = "JHEP",
    volume = "12",
    pages = "071",
    year = "2015"
}

@article{Komargodski:2026ain,
    author = "Komargodski, Zohar and Miscioscia, Alessio and Popov, Fedor K.",
    title = "{Regge's Inferno}",
    eprint = "2603.10197",
    archivePrefix = "arXiv",
    primaryClass = "hep-th",
    month = "3",
    year = "2026"
}

@article{Monin:2016jmo,
    author = "Monin, Alexander and Pirtskhalava, David and Rattazzi, Riccardo and Seibold, Fiona K.",
    title = "{Semiclassics, Goldstone Bosons and CFT data}",
    eprint = "1611.02912",
    archivePrefix = "arXiv",
    primaryClass = "hep-th",
    doi = "10.1007/JHEP06(2017)011",
    journal = "JHEP",
    volume = "06",
    pages = "011",
    year = "2017"
}

@article{Cuomo:2021cnb,
    author = "Cuomo, Gabriel and Mezei, M{\'a}rk and Raviv-Moshe, Avia",
    title = "{Boundary conformal field theory at large charge}",
    eprint = "2108.06579",
    archivePrefix = "arXiv",
    primaryClass = "hep-th",
    doi = "10.1007/JHEP10(2021)143",
    journal = "JHEP",
    volume = "10",
    pages = "143",
    year = "2021"
}

@article{Shaghoulian:2015kta,
    author = "Shaghoulian, Edgar",
    title = "{Modular forms and a generalized Cardy formula in higher dimensions}",
    eprint = "1508.02728",
    archivePrefix = "arXiv",
    primaryClass = "hep-th",
    doi = "10.1103/PhysRevD.93.126005",
    journal = "Phys. Rev. D",
    volume = "93",
    number = "12",
    pages = "126005",
    year = "2016"
}

@article{Simmons-Duffin:2016wlq,
    author = "Simmons-Duffin, David",
    title = "{The Lightcone Bootstrap and the Spectrum of the 3d Ising CFT}",
    eprint = "1612.08471",
    archivePrefix = "arXiv",
    primaryClass = "hep-th",
    doi = "10.1007/JHEP03(2017)086",
    journal = "JHEP",
    volume = "03",
    pages = "086",
    year = "2017"
}

@article{Caron-Huot:2017vep,
    author = "Caron-Huot, Simon",
    title = "{Analyticity in Spin in Conformal Theories}",
    eprint = "1703.00278",
    archivePrefix = "arXiv",
    primaryClass = "hep-th",
    doi = "10.1007/JHEP09(2017)078",
    journal = "JHEP",
    volume = "09",
    pages = "078",
    year = "2017"
}

@article{Fardelli:2025fkn,
    author = "Fardelli, Giulia and Fitzpatrick, A. Liam and Li, Wei",
    title = "{Towards Large-Spin Effective Theory II: $O(2)$ model in $d=4-\epsilon$}",
    eprint = "2508.20160",
    archivePrefix = "arXiv",
    primaryClass = "hep-th",
    month = "8",
    year = "2025"
}

@article{Alday:2007mf,
    author = "Alday, Luis F. and Maldacena, Juan Martin",
    title = "{Comments on operators with large spin}",
    eprint = "0708.0672",
    archivePrefix = "arXiv",
    primaryClass = "hep-th",
    doi = "10.1088/1126-6708/2007/11/019",
    journal = "JHEP",
    volume = "11",
    pages = "019",
    year = "2007"
}

@article{Deb:2025ddc,
    author = "Deb, Anirudh",
    title = "{Generalized Schur limit, modular differential equations and quantum monodromy traces}",
    eprint = "2512.02102",
    archivePrefix = "arXiv",
    primaryClass = "hep-th",
    reportNumber = "YITP-SB-2025-20",
    month = "12",
    year = "2025"
}

@article{Deb:2025ypl,
    author = "Deb, Anirudh and Razamat, Shlomo S.",
    title = "{Generalized Schur partition functions and RG flows}",
    eprint = "2506.13764",
    archivePrefix = "arXiv",
    primaryClass = "hep-th",
    reportNumber = "YITP-SB-2025-12",
    doi = "10.1103/ldxd-2jm5",
    journal = "Phys. Rev. D",
    volume = "113",
    number = "4",
    pages = "045011",
    year = "2026"
}

@article{Banerjee:2012iz,
    author = "Banerjee, Nabamita and Bhattacharya, Jyotirmoy and Bhattacharyya, Sayantani and Jain, Sachin and Minwalla, Shiraz and Sharma, Tarun",
    title = "{Constraints on Fluid Dynamics from Equilibrium Partition Functions}",
    eprint = "1203.3544",
    archivePrefix = "arXiv",
    primaryClass = "hep-th",
    reportNumber = "TFR-TH-12-05, IPMU12-0037",
    doi = "10.1007/JHEP09(2012)046",
    journal = "JHEP",
    volume = "09",
    pages = "046",
    year = "2012"
}

@article{Kravchuk:2024wmv,
    author = "Kravchuk, Petr and Mann, Jeremy A.",
    title = "{AdS N-body problem at large spin}",
    eprint = "2412.12328",
    archivePrefix = "arXiv",
    primaryClass = "hep-th",
    doi = "10.1007/JHEP10(2025)004",
    journal = "JHEP",
    volume = "10",
    pages = "004",
    year = "2025"
}

@article{Cuomo:2022kio,
    author = "Cuomo, Gabriel and Komargodski, Zohar",
    title = "{Giant Vortices and the Regge Limit}",
    eprint = "2210.15694",
    archivePrefix = "arXiv",
    primaryClass = "hep-th",
    doi = "10.1007/JHEP01(2023)006",
    journal = "JHEP",
    volume = "01",
    pages = "006",
    year = "2023"
}

@article{Barrat:2025wbi,
    author = "Barrat, Julien and Marchetto, Enrico and Miscioscia, Alessio and Pomoni, Elli",
    title = "{Thermal Bootstrap for the Critical O(N) Model}",
    eprint = "2411.00978",
    archivePrefix = "arXiv",
    primaryClass = "hep-th",
    reportNumber = "DESY-24-167",
    doi = "10.1103/PhysRevLett.134.211604",
    journal = "Phys. Rev. Lett.",
    volume = "134",
    number = "21",
    pages = "211604",
    year = "2025"
}

@article{Metsaev:2002re,
    author = "Metsaev, R. R. and Tseytlin, Arkady A.",
    title = "{Exactly solvable model of superstring in Ramond-Ramond plane wave background}",
    eprint = "hep-th/0202109",
    archivePrefix = "arXiv",
    reportNumber = "FIAN-TD-02-04",
    doi = "10.1103/PhysRevD.65.126004",
    journal = "Phys. Rev. D",
    volume = "65",
    pages = "126004",
    year = "2002"
}

@article{mishra2026semiuniversalityonmodels1times,
    author        = {Mishra, Vinayak},
    title         = {{Semi-Universality of $O(N)$ model on $S^1\times S^2$}},
    year          = {2026},
    month         = {9},
    eprint        = {2609.25622},
    archivePrefix = {arXiv},
    primaryClass  = {hep-th},
    url           = {https://arxiv.org/abs/2609.25622}
}

@article{Sachdev:1993polylog,
    author        = {Sachdev, Subir},
    title         = {{Polylogarithm identities in a conformal field theory in three dimensions}},
    journal       = {Phys. Lett. B},
    volume        = {309},
    pages         = {285--288},
    year          = {1993},
    doi           = {10.1016/0370-2693(93)90935-B},
    eprint        = {hep-th/9305131},
    archivePrefix = {arXiv}
}

@article{Barrat:2025nvu,
    author = "Barrat, Julien and Bozkurt, Deniz N. and Marchetto, Enrico and Miscioscia, Alessio and Pomoni, Elli",
    title = "{The analytic bootstrap at finite temperature}",
    eprint = "2506.06422",
    archivePrefix = "arXiv",
    primaryClass = "hep-th",
    reportNumber = "DESY-25-078",
    doi = "10.1007/JHEP05(2026)104",
    journal = "JHEP",
    volume = "05",
    pages = "104",
    year = "2026"
}

@article{Diatlyk:2024qpr,
	title        = {{Effective Field Theory of Conformal Boundaries}},
	author       = {Diatlyk, Oleksandr and Khanchandani, Himanshu and Popov, Fedor K. and Wang, Yifan},
	eprint       = {2406.01550},
	archiveprefix = {arXiv},
	primaryclass = {hep-th},
    journal = {Phys. Rev. Lett.},
volume  = {133},
pages   = {261601},
year    = {2024},
doi     = {10.1103/PhysRevLett.133.261601},
}

@article{Simmons-Duffin:2025qox,
    author = "Simmons-Duffin, David and Xu, Yixin",
    title = "{A genus-2 crossing equation in $d\geq 2$}",
    eprint = "2511.07569",
    archivePrefix = "arXiv",
    primaryClass = "hep-th",
    month = "11",
    year = "2025"
}

@article{Mouland:2023gcp,
    author = "Mouland, Rishi",
    title = "{How to build a black hole out of instantons}",
    eprint = "2311.13636",
    archivePrefix = "arXiv",
    primaryClass = "hep-th",
    doi = "10.1007/JHEP03(2024)002",
    journal = "JHEP",
    volume = "03",
    pages = "002",
    year = "2024"
}

@article{Dorey:2023jfw,
    author = "Dorey, Nick and Mouland, Rishi",
    title = "{Conformal quantum mechanics, holomorphic factorisation, and ultra-spinning black holes}",
    eprint = "2302.14850",
    archivePrefix = "arXiv",
    primaryClass = "hep-th",
    doi = "10.1007/JHEP02(2024)086",
    journal = "JHEP",
    volume = "02",
    pages = "086",
    year = "2024"
}

@article{Dorey:2022cfn,
    author = "Dorey, Nick and Mouland, Rishi and Zhao, Boan",
    title = "{Black hole entropy from quantum mechanics}",
    eprint = "2207.12477",
    archivePrefix = "arXiv",
    primaryClass = "hep-th",
    doi = "10.1007/JHEP06(2023)166",
    journal = "JHEP",
    volume = "06",
    pages = "166",
    year = "2023"
}

@article{Lee:2026azy,
    author = "Lee, Eunwoo",
    title = "{Equilibrium Partition Function of Non-Relativistic CFTs in Harmonic Trap}",
    eprint = "2603.09856",
    archivePrefix = "arXiv",
    primaryClass = "hep-th",
    month = "3",
    year = "2026"
}

@article{Mukherjee:2026dfu,
    author = "Mukherjee, Jyotirmoy and Ray, Pabitra",
    title = "{Semi-universality of conformal higher-derivative and conformal higher-spin fields}",
    eprint = "2606.08461",
    archivePrefix = "arXiv",
    primaryClass = "hep-th",
    doi = "10.1007/JHEP08(2026)182",
    journal = "JHEP",
    volume = "08",
    pages = "182",
    year = "2026"
}

@article{Advant:2026cqt,
    author = "Advant, Arnav and Anand, Harsh and Benjamin, Nathan and Kumar, Vipul and Minwalla, Shiraz and Mukherjee, Jyotirmoy and Pal, Sridip and Rahaman, Asikur and Ray, Pabitra",
    title = "{From $Z$ to $a$: High-temperature relations, subleading semi-universality, and conformal anomalies}",
    eprint = "2609.03013",
    archivePrefix = "arXiv",
    primaryClass = "hep-th",
    reportNumber = "TIFR/TH/26-24",
    month = "9",
    year = "2026"
}

@article{Maldacena:2008w,
    author = "Maldacena, Juan and Martelli, Dario and Tachikawa, Yuji",
    title = "{Comments on string theory backgrounds with non-relativistic conformal symmetry}",
    eprint = "0807.1100",
    archivePrefix = "arXiv",
    primaryClass = "hep-th",
    doi = "10.1088/1126-6708/2008/10/072",
    journal = "JHEP",
    volume = "10",
    pages = "072",
    year = "2008"
}

@article{Buric:2026pes,
    author = "Buri{\'c}, Ilija and Mangialardi, Francesco and Russo, Francesco and Schomerus, Volker and Vichi, Alessandro",
    title = "{Thermal One-point Functions and Asymptotic CFT Data: QFT in AdS}",
    eprint = "2606.17167",
    archivePrefix = "arXiv",
    primaryClass = "hep-th",
    month = "6",
    year = "2026"
}

@article{Fitzpatrick:2012yx,
	title        = {{The Analytic Bootstrap and AdS Superhorizon Locality}},
	author       = {Fitzpatrick, A. Liam and Kaplan, Jared and Poland, David and Simmons-Duffin, David},
	year         = 2013,
	journal      = {JHEP},
	volume       = 12,
	pages        = {004},
	doi          = {10.1007/JHEP12(2013)004},
	eprint       = {1212.3616},
	archiveprefix = {arXiv},
	primaryclass = {hep-th}
}

@article{Poland:2018epd,
	title        = {{The Conformal Bootstrap: Theory, Numerical Techniques, and Applications}},
	author       = {Poland, David and Rychkov, Slava and Vichi, Alessandro},
	year         = 2019,
	journal      = {Rev. Mod. Phys.},
	volume       = 91,
	pages        = {015002},
	doi          = {10.1103/RevModPhys.91.015002},
	eprint       = {1805.04405},
	archiveprefix = {arXiv},
	primaryclass = {hep-th}
}

@article{Benjamin:2023qsc,
    author = "Benjamin, Nathan and Lee, Jaeha and Ooguri, Hirosi and Simmons-Duffin, David",
    title = "{Universal asymptotics for high energy CFT data}",
    eprint = "2306.08031",
    archivePrefix = "arXiv",
    primaryClass = "hep-th",
    reportNumber = "CALT-TH 2023-014, IPMU 23-0020",
    doi = "10.1007/JHEP03(2024)115",
    journal = "JHEP",
    volume = "03",
    pages = "115",
    year = "2024"
}

@article{Alday:2015eya,
    author = "Alday, Luis F. and Bissi, Agnese and Lukowski, Tomasz",
    title = "{Large spin systematics in CFT}",
    eprint = "1502.07707",
    archivePrefix = "arXiv",
    primaryClass = "hep-th",
    doi = "10.1007/JHEP11(2015)101",
    journal = "JHEP",
    volume = "11",
    pages = "101",
    year = "2015"
}

@article{Jensen:2011xb,
	title        = {{Parity-Violating Hydrodynamics in 2+1 Dimensions}},
	author       = {Jensen, Kristan and Kaminski, Matthias and Kovtun, Pavel and Meyer, Rene and Ritz, Adam and Yarom, Amos},
	year         = 2012,
	journal      = {JHEP},
	volume       = 05,
	pages        = 102,
	doi          = {10.1007/JHEP05(2012)102},
	eprint       = {1112.4498},
	archiveprefix = {arXiv},
	primaryclass = {hep-th}
}

@article{Manes:2013kka,
	title        = {{Parity odd equilibrium partition function in 2+1 dimensions}},
	author       = {Ma{\~n}es, Juan L. and Valle, Manuel},
	year         = 2013,
	journal      = {JHEP},
	volume       = 11,
	pages        = 178,
	doi          = {10.1007/JHEP11(2013)178},
	eprint       = {1310.2113},
	archiveprefix = {arXiv},
	primaryclass = {hep-th}
}

@article{Komargodski:2012ek,
    author = "Komargodski, Zohar and Zhiboedov, Alexander",
    title = "{Convexity and Liberation at Large Spin}",
    eprint = "1212.4103",
    archivePrefix = "arXiv",
    primaryClass = "hep-th",
    doi = "10.1007/JHEP11(2013)140",
    journal = "JHEP",
    volume = "11",
    pages = "140",
    year = "2013"
}

@article{Fardelli:2025eun,
    author = "Fardelli, Giulia and Fitzpatrick, A. Liam and Li, Wei",
    title = "{Towards Large-Spin Effective Theory I: Three-Particle States in AdS $\phi^4$ Theory}",
    eprint = "2508.20158",
    archivePrefix = "arXiv",
    primaryClass = "hep-th",
    journal = {JHEP},
volume  = {06},
pages   = {004},
year    = {2026},
doi     = {10.1007/JHEP06(2026)004},
}

@article{SemiUniversality,
    author = "Anand, Harsh and Benjamin, Nathan and Kumar, Vipul and Minwalla, Shiraz and Mukherjee, Jyotirmoy and Pal, Sridip and Rahaman, Asikur",
    title = "{Semi-universality of CFT$_d$ entropy at large spin}",
    eprint = "2512.00158",
    archivePrefix = "arXiv",
    primaryClass = "hep-th",
    journal = {SciPost Phys.},
volume  = {21},
number  = {2},
pages   = {042},
year    = {2026},
doi     = {10.21468/SciPostPhys.21.2.042},
    
}

@article{Katz:2016hxp,
    author = "Katz, Emanuel and Khandker, Zuhair U. and Walters, Matthew T.",
    title = "{A Conformal Truncation Framework for Infinite-Volume Dynamics}",
    eprint = "1604.01766",
    archivePrefix = "arXiv",
    primaryClass = "hep-th",
    doi = "10.1007/JHEP07(2016)140",
    journal = "JHEP",
    volume = "07",
    pages = "140",
    year = "2016"
}

@article{Marchetto:2023xap,
    author = "Marchetto, Enrico and Miscioscia, Alessio and Pomoni, Elli",
    title = "{Sum rules \& Tauberian theorems at finite temperature}",
    eprint = "2312.13030",
    archivePrefix = "arXiv",
    primaryClass = "hep-th",
    reportNumber = "DESY-23-224",
    doi = "10.1007/JHEP09(2024)044",
    journal = "JHEP",
    volume = "09",
    pages = "044",
    year = "2024"
}

@article{Iliesiu:2018zlz,
	title        = {{Bootstrapping the 3d Ising model at finite temperature}},
	author       = {Iliesiu, Luca and Kolo\u{g}lu, Murat and Simmons-Duffin, David},
	year         = 2019,
	journal      = {JHEP},
	volume       = 12,
	pages        = {072},
	doi          = {10.1007/JHEP12(2019)072},
	eprint       = {1811.05451},
	archiveprefix = {arXiv},
	primaryclass = {hep-th},
	reportnumber = {CALT-TH-2018-049, PUPT-2573}
}

@article{Petkou:2018ynm,
	title        = {{Dynamics of Finite-Temperature Conformal Field Theories from Operator Product Expansion Inversion Formulas}},
	author       = {Petkou, Anastasios C. and Stergiou, Andreas},
	year         = 2018,
	journal      = {Phys. Rev. Lett.},
	volume       = 121,
	number       = 7,
	pages        = {071602},
	doi          = {10.1103/PhysRevLett.121.071602},
	eprint       = {1806.02340},
	archiveprefix = {arXiv},
	primaryclass = {hep-th},
	reportnumber = {CERN-TH-2018-132}
}

@article{David:2023uya,
	title        = {{Thermal one-point functions: CFT\textquoteright{}s with fermions, large d and large spin}},
	author       = {David, Justin R. and Kumar, Srijan},
	year         = 2023,
	journal      = {JHEP},
	volume       = 10,
	pages        = 143,
	doi          = {10.1007/JHEP10(2023)143},
	eprint       = {2307.14847},
	archiveprefix = {arXiv},
	primaryclass = {hep-th}
}

@article{Cardy:1986ie,
	title        = {{Operator Content of Two-Dimensional Conformally Invariant Theories}},
	author       = {Cardy, John L.},
	year         = 1986,
	journal      = {Nucl. Phys. B},
	volume       = 270,
	pages        = {186--204},
	doi          = {10.1016/0550-3213(86)90552-3}
}

@misc{Blau:PlaneWavesPenroseLimits,
    author = {Blau, Matthias},
    title = {{Plane Waves and Penrose Limits}},
    howpublished = {Lecture notes, November 15, 2011,
        \url{https://ncatlab.org/nlab/files/BlauPlaneWavesAndPenroseLimit.pdf}}
}

@article{BMN2002,
  author  = {Berenstein, David and Maldacena, Juan and Nastase, Horatiu},
  title   = {{Strings in flat space and pp waves from N=4 Super Yang Mills}},
  journal = {JHEP},
  volume  = {04},
  pages   = {013},
  year    = {2002},
  doi     = {10.1088/1126-6708/2002/04/013},
  eprint  = {hep-th/0202021},
  archivePrefix = {arXiv}
}

@article{Benjamin:2024kdg,
	title        = {{Angular fractals in thermal QFT}},
	author       = {Benjamin, Nathan and Lee, Jaeha and Pal, Sridip and Simmons-Duffin, David and Xu, Yixin},
	year         = 2024,
	journal      = {JHEP},
	volume       = 11,
	pages        = 134,
	doi          = {10.1007/JHEP11(2024)134},
	eprint       = {2405.17562},
	archiveprefix = {arXiv},
	primaryclass = {hep-th},
	reportnumber = {CALT-TH 2024-021}
}

@article{vanRees:2024xkb,
    author = "van Rees, Balt C.",
    title = "{Theorems for the lightcone bootstrap}",
    eprint = "2412.06907",
    archivePrefix = "arXiv",
    primaryClass = "hep-th",
    doi = "10.21468/SciPostPhys.18.6.207",
    journal = "SciPost Phys.",
    volume = "18",
    number = "6",
    pages = "207",
    year = "2025"
}

@article{Sachdev:1992py,
	title        = {{Universal quantum critical dynamics of two-dimensional antiferromagnets}},
	author       = {Sachdev, Subir and Ye, Jinwu},
	year         = 1992,
	journal      = {Phys. Rev. Lett.},
	volume       = 69,
	pages        = 2411,
	doi          = {10.1103/PhysRevLett.69.2411},
	eprint       = {cond-mat/9204001},
	archiveprefix = {arXiv},
	reportnumber = {NSF-ITP-92-54}
}

@article{David:2026yis,
    author = "David, Justin R. and Kumar, Srijan",
    title = "{The large $N$ vector model with angular velocity}",
    eprint = "2608.31151",
    archivePrefix = "arXiv",
    primaryClass = "hep-th",
    reportNumber = "ICTS-USTC/PCFT-26-58",
    month = "8",
    year = "2026"
}

@article{David:2025tqn,
    author = "David, Justin R. and Kumar, Srijan",
    title = "{High to low temperature: O(N) model at large N}",
    eprint = "2508.14872",
    archivePrefix = "arXiv",
    primaryClass = "hep-th",
    doi = "10.1007/JHEP02(2026)194",
    journal = "JHEP",
    volume = "02",
    pages = "194",
    year = "2026"
}

@article{Iliesiu:2018fao,
	title        = {{The Conformal Bootstrap at Finite Temperature}},
	author       = {Iliesiu, Luca and Kolo\u{g}lu, Murat and Mahajan, Raghu and Perlmutter, Eric and Simmons-Duffin, David},
	year         = 2018,
	journal      = {JHEP},
	volume       = 10,
	pages        = {070},
	doi          = {10.1007/JHEP10(2018)070},
	eprint       = {1802.10266},
	archiveprefix = {arXiv},
	primaryclass = {hep-th},
	reportnumber = {CALT-TH-2018-013, PUPT-2550}
}

@article{Buric:2025uqt,
    author = "Buri{\'c}, Ilija and Mangialardi, Francesco and Russo, Francesco and Schomerus, Volker and Vichi, Alessandro",
    title = "{Heavy-Heavy-Light Asymptotics from Thermal Correlators}",
    eprint = "2506.21671",
    archivePrefix = "arXiv",
    primaryClass = "hep-th",
    journal = {JHEP},
volume  = {04},
pages   = {027},
year    = {2026},
doi     = {10.1007/JHEP04(2026)027},
}
\bibliographystyle{JHEP}

\end{document}